\documentclass{aa}  
\usepackage{natbib}
\bibpunct{(}{)}{;}{a}{}{,} 
\usepackage{graphicx}

\usepackage{txfonts}
\usepackage{lscape}
\usepackage[breaklinks=true]{hyperref}
\hypersetup{colorlinks,linkcolor=red,citecolor=blue,urlcolor=blue}

\newcommand{\Teff}{$T_{\rm eff}$}
\newcommand{\logg}{$\log g$}
\newcommand{\FeH}{[Fe/H]}
\newcommand{\ubv}{\ensuremath{UB\,V}}
\newcommand{\rgc}{$R_{\rm GC}$}
\usepackage{amstext}
\usepackage{color}

\begin{document}

   \title{On the metallicity of open clusters. III. Homogenised sample}


   \author{M. Netopil
          \inst{1}
          \and
          E. Paunzen\inst{1}
          \and
          U. Heiter\inst{2}
          \and
          C. Soubiran\inst{3}
          }

   \institute{Department of Theoretical Physics and Astrophysics, Masaryk University, Kotl\'a\v{r}sk\'a 2, 611 37 Brno, Czech Republic \\
             \email{mnetopil@physics.muni.cz}
         \and
         Institutionen för fysik och astronomi, Uppsala universitet, Box 516, 751 20 Uppsala, Sweden
         \and
         Universit\'e de Bordeaux - CNRS, LAB - UMR 5804, BP 89, 33270 Floirac, France
}

   \date{Received <date>; accepted <date>}

 
  \abstract
   {Open clusters are known as excellent tools for various topics in Galactic research. For example, they allow accurately tracing the chemical structure of the Galactic disc. However, the metallicity is known only for a rather low percentage of the open cluster population, and these values are based on a variety of methods and data. Therefore, a large and homogeneous sample is highly desirable.}
   {In the third part of our series we compile a large sample of homogenised open cluster metallicities using a wide variety of different sources. These data and a sample of Cepheids are used to investigate the radial metallicity gradient, age effects, and to test current models.}
   {We used photometric and spectroscopic data to derive cluster metallicities. The different sources were checked and tested for possible offsets and correlations.}
   {In total, metallicities for 172 open cluster were derived. We used the spectroscopic data of 100 objects for a study of the radial metallicity distribution and the age-metallicity relation. We found a possible increase of metallicity with age, which, if confirmed, would provide observational evidence for radial migration. Although a statistical significance is given, more studies are certainly needed to exclude selection effects, for example. The comparison of open clusters and Cepheids with recent Galactic models agrees well in general. However, the models do not reproduce the flat gradient of the open clusters in the outer disc. Thus, the effect of radial migration is either underestimated in the models, or an
additional mechanism is at work.}
   {Apart from the Cepheids, open clusters are the best tracers for metallicity over large Galactocentric distances in the Milky Way. For a sound statistical analysis, a sufficiently large and homogeneous sample of cluster metallicities is needed. Our compilation is currently by far the largest and provides the basis for several basic studies such as the statistical treatment of the Galactic cluster population, or evolutionary studies of individual star groups in open clusters.}

   \keywords{Galaxy: abundances –- Galaxy: structure –- open clusters and associations: general }

   \maketitle
%

\section{Introduction}

The metallicity of open clusters is by far the least well-known parameter, but is a key for a more profound understanding
of stellar formation and evolution or for investigating the chemical properties of our Galaxy. The updated open cluster catalogue by \citet[][version 3.4]{dias02} lists metallicity estimates for about 280 open clusters, which represent about 13 percent of the known cluster population in the Milky Way. The results were adopted from many different sources and are based on spectroscopic data of different quality or on various photometric methods (e.g. metallicity-dependent indices or isochrone fitting). The listed metallicities therefore cannot be considered as a homogeneous data set. Sufficiently large samples of open cluster metallicities are needed, and they need to have been calibrated in a consistent way with a single method. Only samples like this allow a proper comparison and identifying possible offsets or trends in the data. 

To our knowledge, \citet{gratton00} was the only one who homogenised metallicities of open clusters using estimates based on more than two different methods. His final list consists of 104 open clusters. By now, many more datasets are available, in particular based on high-resolution spectra that are expected to provide the most accurate way to derive the iron abundance and allow a better revision of the various metallicity scales. We wish to mention the ongoing \textit{Gaia}-ESO Public Spectroscopic Survey \citep{gilmore12}, which has already provided mean abundances for some open clusters.

In our series we investigate the current status of open cluster metallicities. The first paper \citep[][Paper I]{paunzen10} was dedicated to a compilation of photometric results, and in the second part \citep[][Paper II]{heiter14} we provided mean metallicities based on high-resolution, high S/N spectroscopic literature data. This third paper combines the two previous studies and provides homogenised metallicity values for a large sample of open clusters. 

We present evaluated metallicities for 172 open clusters in total that we combine with metallicities of Cepheids to study the distribution of the iron abundance in the Milky Way. Furthermore, these two samples are compared with recent chemical
evolution models. Throughout this article, we use the term metallicity synonymously with iron abundance [Fe/H]. 

The article is arranged as follows: in Sect. \ref{sect:sources} we describe the
selection of the metallicity determinations. In Sect. \ref{sect:metcomp} we compare the individual results and provide a final sample with recalibrated metallicities. In Sect. \ref{sect:results} we discuss the Galactic distribution of the sample, study dependencies on the age, and compare the data with recent chemical models. Finally, Sect. \ref{sect:conclusion} concludes the paper.

\section{Data sources for open clusters}
\label{sect:sources}
\subsection{Spectroscopic metallicities}

In Paper II we have evaluated available spectroscopic iron abundance determinations of open cluster stars and presented mean values for 78 open clusters. The results are based on high-resolution data ($R \geq 25000$) with high signal-to-noise ratios ($S/N \geq 50$). Furthermore, quality criteria were introduced by adopting only \FeH\ measurements of stars with \Teff\ $=$ 4400–6500\,K and \logg\ $\geq$ 2.0\,dex. We have to note that the mean iron abundance for Berkeley~29, listed in Paper II, also incorporates some measurements based upon lower $S/N$ data. In the present paper we therefore list the correct values for the higher and lower quality data. Since publication of Paper II, some new studies
were made  \citep{boesg13,bocek15,carraro14b,donati15,magrini14,magrini15,molenda14,monaco14,reddy15}, which we examined the same way as described in Paper II. This adds ten open clusters to our list (Berkeley~81, NGC~1342, NGC~1662, NGC~1912, NGC~2354 NGC~4337, NGC~4815, NGC~6811,  Trumpler~5, and Trumpler~20) and supplementary data for NGC~752, NGC~2447, NGC~2632, and NGC~6705.

These mean cluster metallicities based on high-quality spectroscopic data (HQS) serve here as primary standards to verify and calibrate other metallicity determinations, such as photometric ones. As a kind of secondary standard, we combine results based on lower quality spectroscopic (LQS) material (lower in resolution or $S/N$), but again adopting the defined temperature and gravity limits. However, only data with a $S/N$ of at least 20 were considered. In Paper II most of these references are listed, except for the works  on the clusters Melotte~25, NGC~752, NGC~3680, NGC~6253, NGC~6819, and NGC~7789  \citep{atwarog09,twarog10,mader13,mader15,leebrown15,overb15}. Additionally, we included the high-resolution but low $S/N$ data for the clusters Berkeley~81, Melotte~66, NGC~6791, Trumpler~5, and Trumpler~20 \citep{boesg15,brag14,carraro14c,carraro14,magrini15,monaco14}. We have not considered the study on NGC~188 by \citet{randich03} because they added spectra of different stars to obtain a sufficient high $S/N$ for the analysis. Furthermore, we excluded the object NGC~6882/5 that shows three star accumulations at distances between 0.3 to 1\,kpc \citep{pena08}. A single star in the cluster area was compiled from the LQS study by \citet{luck94}. However, the parallax as well as the spectral type indicate a much closer distance ($<$\,100\,pc) for the star and hence a non-membership to the cluster(s).

The final spectroscopic sample comprises 100 open clusters, 88 entries based on HQS results, 44 with LQS data, and an overlap of 32 clusters.

\subsection{Photometric metallicities}

In Paper I we have compiled open cluster metallicities based on various photometric systems and calibrations. With the mean spectroscopic metallicity values presented in this work and Paper II we now have a sufficient large sample at our disposal to evaluate the photometric determinations on the system level or even for some individual publications. However, to identify systematic differences, a sufficient large overlap with spectroscopic values is needed. In the following we discuss the available systems and calibrations.

There are several metallicity calibrations available for the DDO system \citep[e.g.][]{janes75,piatti93,twarog96}. These were systematically applied to the same photometric data \citep[e.g.][]{piatti95,twarog97}. Paper I therefore lists in this system in particular several determinations for the same cluster that might influence the mean values listed in this paper. To avoid such a bias, we intend to select one determination in each photometric system. This is in general the latest available study because with the continuously increasing amount of different data, a more detailed  membership analysis, for example, is expected. For the DDO system we adopted the work by \citet{twarog97}, who re-analysed all so far available data for about 60 open clusters. The results of more recent studies were adopted in a few cases because  \citet{claria06,claria08}\footnote{We
note that the first reference is missing in Paper I.} , for instance,
presented new photometry, but also a detailed membership analysis based on CORAVEL radial-velocity observations. These metallicity determinations are all on the scale by \citet{piatti95}, which we have transformed to the scale used by \citet{twarog97} with the corrections given by the latter reference. The  list of DDO metallicity estimates consists of 67 open clusters.

The metallicity studies using the Str\"omgren line blanketing indicator $m_1$ \citep{stroemgren66} are based upon the calibrations by \citet{nissen81,nissen88}. They only differ by an offset of 0.04\,dex owing to the adopted zero-points of the Hyades metallicity. We have scaled all results to the later calibration, which uses an iron abundance of 0.12\,dex in agreement to modern spectroscopic values. In case of multiple studies on the same cluster, we have again chosen the most recent determination. The results for the clusters NGC~6192 and NGC~6451 \citep{paunzen03} were rejected because the CCD photometry appears to be insufficiently standardised \citep[see discussion by][]{netopil13}. Some open cluster results were added to the list in Paper I: NGC~5822, NGC~6253, NGC~6791, and NGC~6819 \citep{atwarog14,carraro11,twarog07}, resulting in 28 open clusters in total. 

The last three references provide supplementary measurements in a filter that covers the H and K lines of Ca\,{\sc ii}, as part of the $Caby$ system \citep{twarog91}. With this system another metallicity-dependent index can be derived ($\delta hk$). So far, only nine open clusters were studied, but the objects cover a broad metallicity range ($\sim$\,1\,dex). Although the calibration is strongly tied to the $m_1$ calibration by \citet{nissen88}, we treat them as independent metallicity estimates. 

The probably most often used photometric system in open cluster research is the \ubv\ system, for which several metallicity calibrations available \citep[e.g.][]{carney79,cameron85a,karschu06,karaali11}. However, most of them are little used, except the calibration by \citet{cameron85a}. Two studies are available \citep{cameron85b,tadross03} that present \ubv\ metallicity estimates for 38 and 91 clusters, respectively. The main apparent difference between these two studies is that \citet{cameron85b} used photoelectric data ($\ubv_p$), while \citet{tadross03} applied the calibration to CCD data ($\ubv_c$). From the $\ubv_c$ sample we excluded Berkeley~42, which is a globular cluster \citep[][]{harris96}. However, with 123 open clusters in total (there is an overlap of only five objects), the \ubv\ system provides by far the largest number of open cluster metallicities.

\begin{figure*}
\centering
\includegraphics[width=150mm]{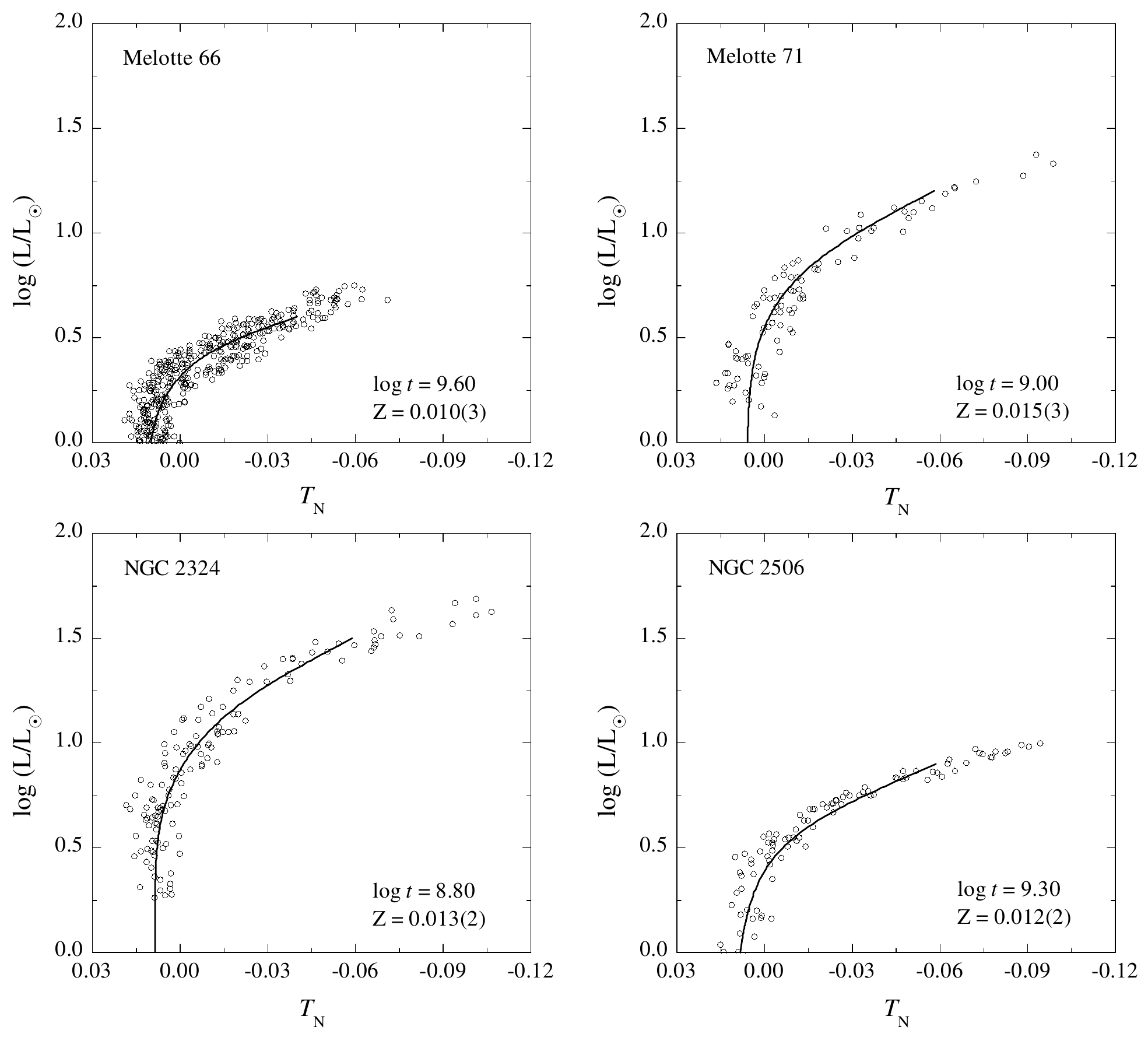}
\caption{Results for the four additional open clusters analysed using the method by \citet{poehnl10}. The lines represent the best-fitting isochrones, and $T_N$ is the temperature difference in dex between the star and the ZAMS at solar metallicity. } 
\label{fig:grids}
\end{figure*}

Another method to determine metallicity was introduced by \citet{poehnl10}. It relies on photometric data and evolutionary models by using zero-age main-sequence (ZAMS) normalised isochrones (differential grids, DG). The advantages of this method are that it is almost independent of the photometric system and that the whole cluster main sequence can be used to determine the metallicity, but all other cluster parameters such as the reddening, distance, and the age can be derived as well. Recently, \citet{netopil13} analysed about 60 open clusters with this method, resulting in an excellent agreement with spectroscopic determinations (see also discussion in Paper II). However, a better coverage at the metal-poor end is desirable to further verify the scale of this method. We therefore selected four open clusters with sufficient photometric data in WEBDA\footnote{http://webda.physics.muni.cz} and for which a subsolar metallicity is listed in Paper II. These are Melotte~66, Melotte~71, NGC~2324, and NGC~2506 with iron abundances  \FeH\ $< -0.2$\,dex. The analysis was performed as described by \citet{netopil13} using photometric data listed in WEBDA. The results of the fitting process can be found in Fig. \ref{fig:grids} and Table \ref{tab:griddata}. Including these new results, the one for Collinder~261 (Paper II), and the clusters already analysed by \citet{poehnl10} and \citet{netopil13}, metallicity estimates for 70 open clusters are available. We adopted the results by the latter reference for objects in common with \citet{poehnl10} because they are based on a broader selection of photometric data.

\begin{table*}
\caption{Derived parameters for the four additional open clusters.} 
\label{tab:griddata} 
\centering 
\begin{tabular}{l l c c l l l} 
\hline\hline 
Cluster & $\log t$ & $(m-M)_{0}$ & $E(B-V)$ & Z$_m$\tablefootmark{a,b} & \FeH\ \tablefootmark{b,c} & phot. systems\tablefootmark{d}\\ 
\hline 
Melotte~66 & 9.60 & 13.35 & 0.14 & 0.010(3) & $-$0.29(14) & UBVI$_{\rm ccd}$, Str$_{\rm ccd}$\\
Melotte~71 & 9.00 & 11.60 & 0.19 & 0.015(3) & $-$0.13(10)  & UBVI$_{\rm ccd}$, Str$_{\rm ccd}$\\
NGC~2324 & 8.80 & 13.05 & 0.13 & 0.013(2) & $-$0.20(9)  & UBVI$_{\rm ccd}$\\
NGC~2506 & 9.30 & 12.50 & 0.04 & 0.012(2) & $-$0.23(9)  & UBV$_{\rm pe}$, UBVRI$_{\rm ccd}$\\
\hline 
\end{tabular}
\tablefoot{
\tablefoottext{a}{We use the subscript ``m'' to distinguish the mass fraction of elements other than H and He from Galactic XYZ coordinates.}
\tablefoottext{b}{The errors of the last significant digits are given in parentheses.}
\tablefoottext{c}{[Fe/H] calculated from Z$_m$ according to \citet{poehnl10}.}
\tablefoottext{d}{Photometric systems used for the analysis: e.g. Str (Str\"omgren), pe/ccd stands for photoelectric and CCD data, respectively.
}
}
\end{table*}

We considered two additional methods in the photometric category as well, although they are based on (low-resolution) spectroscopic data. \citet{friel02} presented an update of previous investigations by using spectrophotometric indices to derive metallicities of 39 open clusters. Furthermore, from the studies by \citet{carrera07},  \citet{carrera12}, \citet{carrera15}, and \citet{warren09}, we compiled metallicity estimates of 28 open clusters. These authors made use of reduced equivalent widths ($W'$) of the Ca {\sc ii} triplet lines, which show a linear relation with metallicity. Previously, this method was mainly used to investigate globular cluster systems \citep[e.g.][]{rutledge97}. We note that the tabulated \FeH\ values by \citet{warren09} are on a different scale than those by the other two references because they used the results by \citet{friel02} for the metal-rich regime to calibrate $W'$. However, \citet{warren09} provided a calibration based on high-resolution spectroscopic data for the open clusters as well, which we applied to their listed $W'$ values. Hence, all clusters can be assumed to be on the same scale. We excluded the clusters Berkeley~32, NGC~2420, and NGC~2506 from the list by \citet{carrera07} for the reasons described in their paper. 

We did not consider results based on the Washington photometric system \citep{canterna76} because of the strong reddening dependency of the derived metallicities in this system. While for example the DDO metallicities are affected by only 0.01\,dex per 0.01\,mag change of the reddening \citep{twarog96}, the Washington system shows an up to three times higher sensitivity \citep[see][]{geisler91}. A proper revision of the reddening values for the red giant cluster members and a recalculation of the metallicities would be a valuable task, but is beyond the scope of this paper. However, spectroscopic data are already available for two thirds of the about 50 open clusters with metallicity estimates in the Washington system, and there are only about five objects with Washington data as a single photometric metallicity source. Thus, the number of additional objects is quite small. We note that there are also some other recent photometric approaches \citep[e.g.][]{oliveira13,oralhan15}, but so far the overlap with our sample is too small for a proper comparison. 

There are numerous open clusters whose photometric metallicity is based on fewer than five stars (in particular in the DDO and Friel samples). The influence of non-member stars on the results is therefore not negligible. Hence, we checked the membership for stars in about 50 open clusters, where the metallicities rely on that small number of objects. The vast majority of stars are included in the radial velocity survey by \citet{Merm08}. For the few remaining objects we used the additional literature \citep[e.g. proper motion membership by][]{Dias14}. The membership analysis removes four open clusters from the DDO sample (Haffner~8, NGC~2547, NGC~6259, and Pismis~4) and identifies one apparent non-member in the DDO cluster NGC~2972 (star Webda \#2). We recalculated the metallicity based on the remaining two stars with the reddening and calibration used by \citet{twarog97}. We also noticed a typo in the star list for Stock~2 by \citet{Claria96}, HD~13209 is actually HD~13207 and a radial velocity member of the cluster.

The photometric starting sample comprises 219 open clusters, 81 of them studied with at least two different photometric methods, and there is an overlap of 81 clusters with the spectroscopic sample.

\section{Final sample}
\label{sect:metcomp}

\begin{figure*}
\centering
\includegraphics[width=170mm]{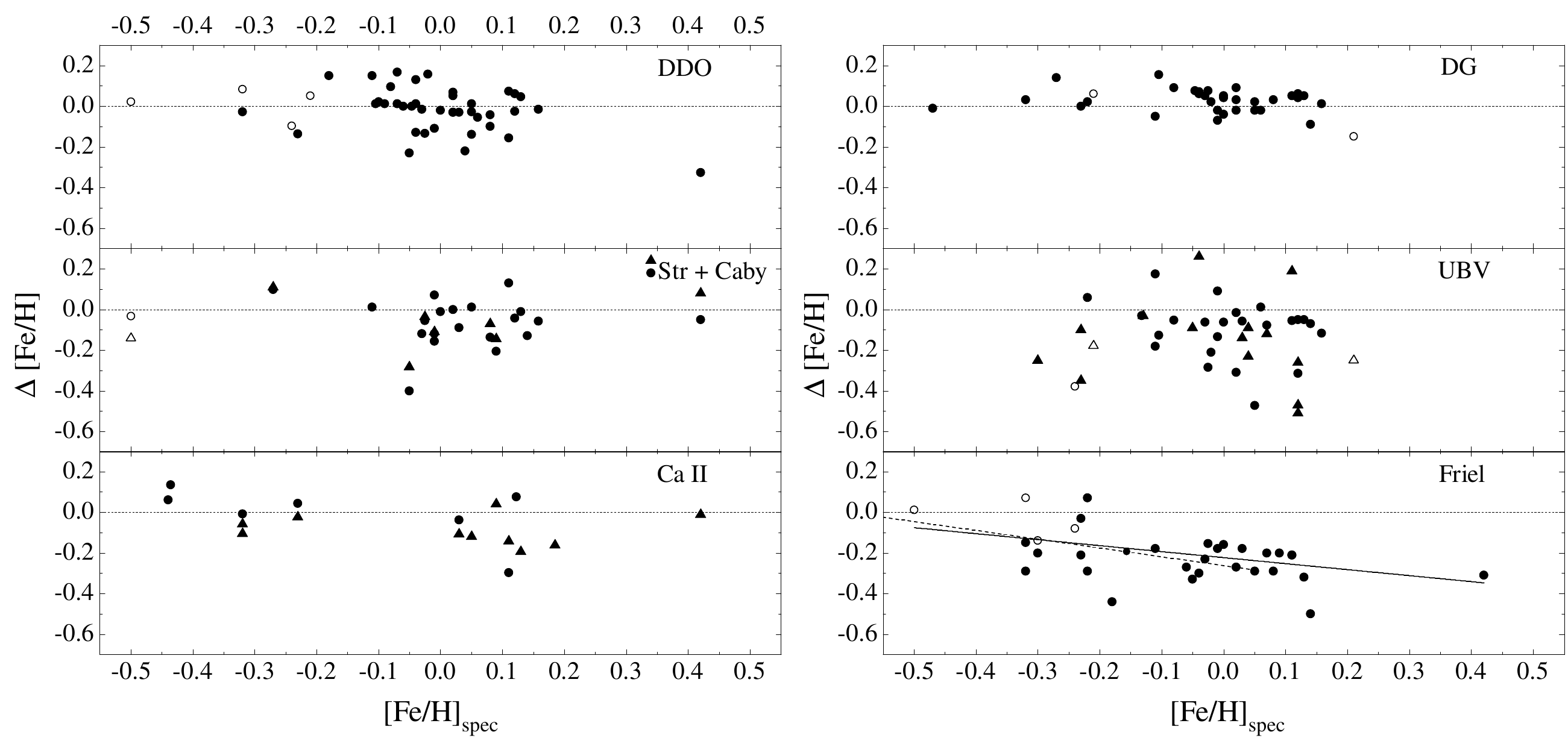}
\caption{Comparison of the various photometric results with the spectroscopic mean values. The differences are given as $\Delta$[Fe/H] = (photometry -- spectroscopy). Filled and open symbols represent HQS and LQS data, respectively. Circles and triangles are used to distinguish Str\"omgren and Caby data, $\ubv_p$ and $\ubv_c$ metallicities, and the Ca{\sc ii} results from the series by Carrera and \citet{warren09}. For the last system there is an overlap of two calibrators (Berkeley~39 and Melotte~66), and both results are shown. For guidance, we always used the same scale. One cluster is outside the range: Berkeley~21 (\ubv$_{c}$, $\Delta$[Fe/H]=$-$0.78). The solid line in the lower right panel shows the linear fit to the Friel data, and the dashed line is the linear fit based on the five calibration clusters as discussed in the text.}
\label{fig:comp}
\end{figure*}

We used the spectroscopic iron abundance values to verify and recalibrate the photometric open cluster results. The comparison is shown in Fig. \ref{fig:comp}. For all systems but one, a constant offset seems justified. Using 2$\sigma$ clipping, we finally derived the corrections with respect to HQS/LQS data (Table \ref{tab:metscales}).

\begin{table}
\caption{Photometric metallicity scales} 
\label{tab:metscales} 
\centering 
\begin{tabular}{l l c c} 
\hline\hline 
System & $\Delta$[Fe/H]\tablefootmark{a} & median & No. of clusters\tablefootmark{b}\\ 
\hline 
DDO & $-$0.001 $\pm$ 0.086 & $-$0.002 & 41/44 \\ 
DG & +0.022 $\pm$ 0.046 & +0.030 & 31/34 \\ 
   & +0.002 $\pm$ 0.045\tablefootmark{c} & +0.010 & 31/34 \\
UBV$_p$ & $-$0.051 $\pm$ 0.089 & $-$0.057 & 20/25 \\ 
UBV$_c$ & $-$0.192 $\pm$ 0.174 & $-$0.180 & 15/17 \\ 
Str\"omgren & $-$0.034 $\pm$ 0.098 & $-$0.043 & 21/22\\
Caby & $-$0.045 $\pm$ 0.104 & $-$0.070 & 7/9\\
Ca {\sc ii} & $-$0.047 $\pm$ 0.098 & $-$0.048 & 14/15 \\
LQS & $-$0.020 $\pm$ 0.089 & $-$0.030 & 31/32 \\
\hline
Friel\tablefootmark{d} & \multicolumn{2}{l}{a=$-$0.223(17); b=$-$0.295(81)} & 28/31 \\
\hline 
\end{tabular}
\tablefoot{
\tablefoottext{a}{The mean differences after 2$\sigma$ clipping are given in the form: photometric system $-$ spectroscopic (HQS/LQS) values, or LQS $-$ HQS.}
\tablefoottext{b}{The number of clusters used to derive the zero-points and the total sample in this category.}
\tablefoottext{c}{The difference if applying the new transformation as discussed in the text.}
\tablefoottext{d}{$\Delta$\FeH = a + b\FeH$_{\rm H/LQS}$; the errors of the last significant digits are given in parenthesis. The mean error after the correction is 0.115\,dex. }}
\end{table}

Most systems provide reasonably scaled results at this point, but two systems differ significantly. In Paper II we briefly discussed that the scale by \citet{friel02}, which is also often denoted as the Boston scale, is lower than the high-resolution spectroscopic results. However, this fact was previouslyy noted by \citet{twarog97} or \citet{gratton00}, who used data from an earlier study \citep{friel93}. Because of the scatter, a constant offset of about $-$0.2\,dex could be justified by using only the HQS data and excluding a few points. However, this might overcalibrate the underabundant clusters. On the other hand, the inclusion of the additional LQS data results in a slope as shown in Fig. \ref{fig:comp} and listed in Table \ref{tab:metscales}. The linear fit is justified after an examination of the values used by \citet{friel02} for their five standard clusters (NGC~2420, NGC~2682, NGC~5904, NGC~6838, and NGC~7789). For the open clusters we adopted the mean abundances derived by us, while for the two globular clusters we used the values listed in the catalogue by \citet[][revision: Dec. 2010]{harris96}. The resulting fit $\Delta$\FeH\ = $-$0.26($\pm$0.07) $-$ 0.43($\pm$0.10)\FeH$_{\rm HQS}$ agrees within the errors with the fit derived before.  

Surprisingly, there are different zero-points for the $\ubv_p$ and $\ubv_c$ data (see Fig. \ref{fig:comp} and Table \ref{tab:metscales}). As already mentioned, different data sources (photoelectric and CCD measurements) were used. Although they are based on the same relation to transform the metallicity indicator $\delta(U-B)_{0.6}$ to \FeH, the deduction of $\delta(U-B)_{0.6}$ differs. While \citet{cameron85b} fitted his two-colour grids to the diagrams of the open clusters, \citet{tadross03} applied the normalisation procedure by \citet{sandage69} to derive a mean $\delta(U-B)_{0.6}$. The $\ubv_c$ data also show the largest scatter, even when excluding the most discrepant results. Unfortunately, neither the number of used cluster stars nor any error estimates are given by \citet{tadross03}. This information is provided by all other compiled references and would be a valuable indicator for the reliability of the results. The colour range $0.1\leq(B-V)_{0}\leq0.7$ is most sensitive to the  ultraviolet excess and was therefore adopted by the two \ubv\ studies. However, we noted that about 20 percent of the $\ubv_c$ clusters are younger than about 15\,Myr. At such a young age only a few main-sequence stars (if any) later than about spectral type A5 can be expected. Another difficulty arises as
a result of differential reddening and a proper exclusion of pre main-sequence objects. However, the results for older clusters appear to be erroneous as well. The data in general show a huge spread in metallicity (up to 1\,dex) over the covered Galactic distance range. We therefore excluded the $\ubv_c$ data set from the final sample, although it is reduced by more than 50 open clusters for which no other metallicity determinations are available. All clusters in this data set have Galactocentric distances \rgc\ $<$ 12\,kpc, thus they do not contribute to the outer Galactic areas at all. It remains an open question whether the applied method or the CCD data themselves are the reason for the apparently faulty results. We note that the standard Johnson \ubv\ system, in particular the $(U-B)$ colour, cannot be reproduced correctly if the instrumental system differs from the originally defined one \citep[see e.g.][]{besssung00,bessell05}. We therefore also excluded the $\ubv_p$ data because the photometric measurements were collected from numerous different sources. For most objects spectroscopic or other probably more reliable photometric results are available, thus the exclusion of these data does not significantly affect the final sample.

Except for the Friel sample, we  generally did not apply the derived corrections given in Table \ref{tab:metscales}. Most of them are too small compared to the scatter and could be a result of the low number of objects involved. We note that even the LQS data show a slight deviation. However, for one system (DG) we applied a correction, motivated as follows. 

The DG method provides well-scaled results, with the lowest scatter of all studied systems. However, the slight overestimation of about 0.02\,dex can be explained by the transformation of Z$_m$ to [Fe/H] values given by \citet{poehnl10}. While they used Z$_m$=0.020 as a solar reference value for the normalisation of the grids, the transformation was based on Z$_m$=0.019 as a solar value (see their Table 2). By applying the transformation [Fe/H] = log(Z$_m$/X$_m$) $-$ log(Z$_m$/X$_m$)$_\sun$ and the values for the grids used by \citet{poehnl10}, the previously found offset vanishes (see Table \ref{tab:metscales}).

Finally, we derived weighted means for the open clusters in the photometric category using in general an equal weight of one. However, because the DG method appears more  accurate compared to the other systems, we decided to adopt a higher weight of 1.5. On the other hand, for individual results based on fewer than five stars we assigned a weight of 0.5 to account for the lower accuracy of single photometric metallicity estimates. 

We rejected the DDO result for NGC~6791 because it deviates strongly compared to other photometric estimates. This object is one of the most metal-rich open clusters with [Fe/H]$\sim$+0.4\,dex; the roughly solar metallicity derived with the DDO system is certainly due to the non-solar scaled element abundances. \citet{carretta07} found considerably underabundant [C/Fe] and [N/Fe] ratios, and the photometric system is sensitive to these elements. These objects can only be unambiguously recognized if several determinations (preferentially spectroscopic analyses) are available. When spectroscopic and photometric results are available, we adopted spectroscopic data for the subsequent analysis. Nevertheless, this clearly demonstrates that individual photometric results can deviate in exceptional cases by up to a few tenths of dex, even if properly reduced and scaled.

Naturally, the spectroscopic results cannot be considered as error free either. The final sample consists of 27 open clusters for which HQS, LQS, and photometric data are available. Among these, we identified six open clusters for which the HQS results deviate by more than 0.1\,dex from both other entries in the same way (Berkeley~21, NGC~188, NGC~2112, NGC~2420, NGC~6253, and NGC~6939). However, by taking into account the individual errors and the number of investigated stars or total photometric weight, the deviations for two open clusters remain significant (NGC~2420 and NGC~6939). We therefore adopted in our analysis the LQS results for these two clusters.

The final sample consist of 172 open clusters, 100 of them with spectroscopic data and 149 for which photometric metallicity determinations are available. This is by far the largest sample of homogenised open cluster metallicities. There are 57 open clusters with at least two photometric measurements, and almost 75 percent of them exhibit a weighted standard deviation lower than 0.1\,dex. The mean cluster metallicities in each category (HQS, LQS, and photometry), the respective errors, and the total photometric weight are listed in Table \ref{tab:longtable}. 

It is beyond the scope of this work to determine the other open cluster parameters (such as the distance and age) by isochrone fitting, for example. We therefore decided to use mean values based on numerous individual studies. This reduces on the one hand the influence of single inaccurate results, but also provides better constraints for the errors. \citet{paunet06} presented a statistical analysis about the accuracy of open cluster parameters. We made use of their parameter compilation and extended them with more recent studies. About 60 additional references provide the parameters for more than 200 open clusters. Furthermore, we included two global parameter studies \citep{buko11,khar13}, both using 2MASS photometry \citep{skrutskie06} and isochrone fitting. The mean cluster parameters (age and distance) of the final open cluster sample are based upon 1437 individual parameter sets. For the majority of the clusters ($\sim$\,90\,\%) at least five determinations were found (median: eight). The error distributions are similar to those presented by \citet{paunet06}, with 80 percent of the objects showing errors in the distance smaller than 20 percent.

\section{Results and discussion}
\label{sect:results}

\subsection{Radial metallicity distribution}
\label{sect:rmd}
\begin{figure}
\centering
\resizebox{\hsize}{!}{\includegraphics{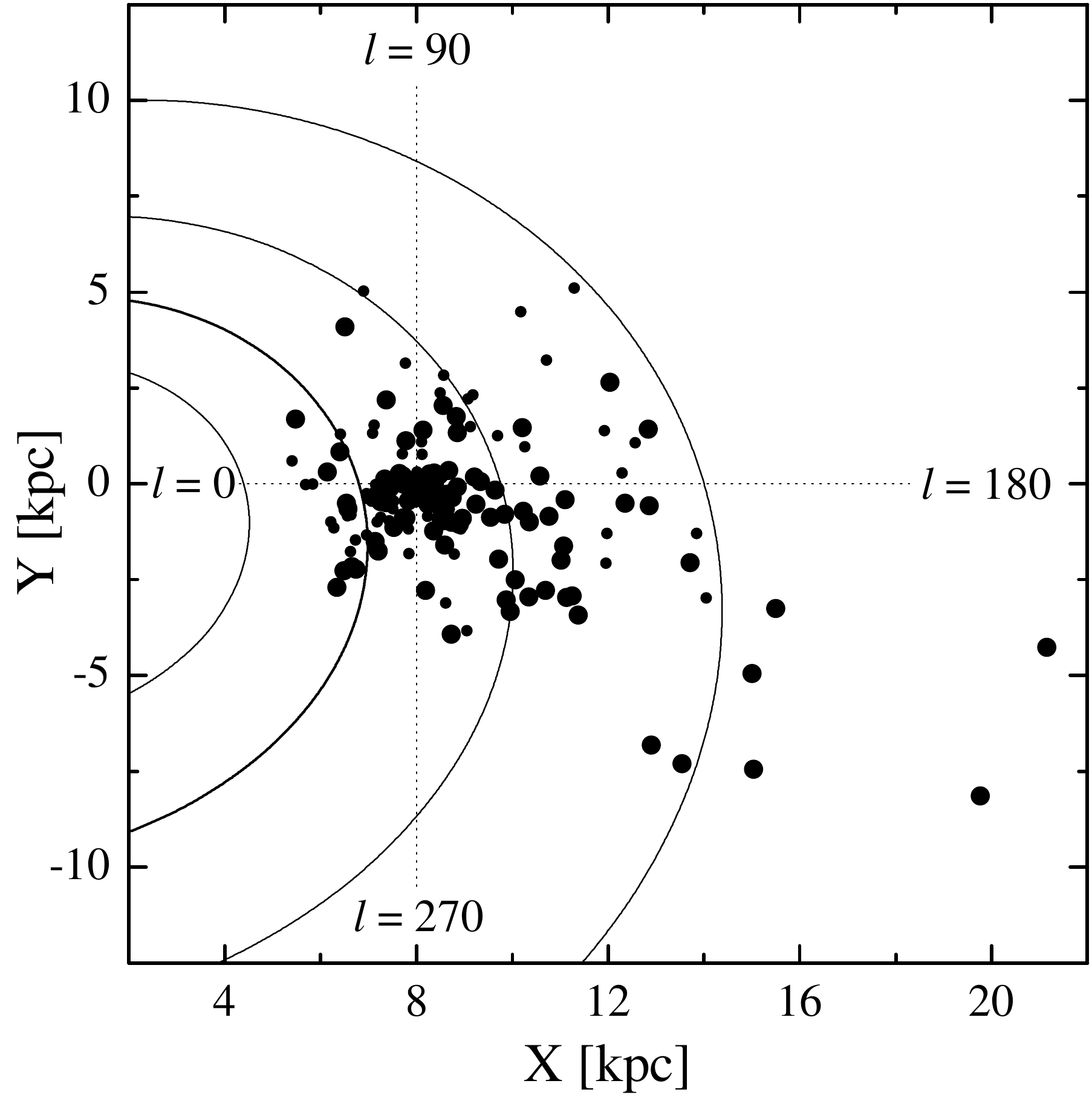}}
\caption{Galactic location of the open clusters projected to the Galactic plane. Open clusters with spectroscopic data are highlighted with larger symbols. The Galactic centre is located at ($X/Y$) = (0/0)\,kpc and the Sun at ($X/Y$) = (8/0)\,kpc. Galactic longitudes in degrees are indicated for better orientation. Furthermore, we show the spiral arm model by \citet{bobylev14} as solid lines. The spiral arms from the inner to the outer area are Scutum-Crux, Carina-Sagittarius, Perseus, and Cygnus.}
\label{fig:xydist}
\end{figure}

Figure \ref{fig:xydist} shows the position of the open clusters projected onto the Galactic plane. As in Paper I and II, we adopted 8\,kpc as the distance of the Sun from the Galactic centre.

With the largest compiled and homogenised sample we are probably able to provide more details for one of the main topics in Galactic research -- the radial metallicity distribution (RMD) of open clusters. Several interpretations of the metallicity gradient in the Milky Way can be found in the literature. The most common results are linear fits over the complete investigated \rgc\ range, as presented for example by \citet{chen03}, \citet{friel02}, and many others, but also gradients with different slopes for inner and outer areas were applied \citep[e.g. by][]{andrievsky04,yong12}. Yong and collaborators determined the transition radius between the inner and outer disc in a quantitative and qualitative manner, resulting in 12\,kpc and 13\,kpc, respectively. While the inner disc shows a steeper gradient, the metallicity distribution of the outer area is almost flat. We note that they adopted the same Galactocentric distance of the Sun as we did. 

\begin{figure}
\centering
\resizebox{\hsize}{!}{\includegraphics{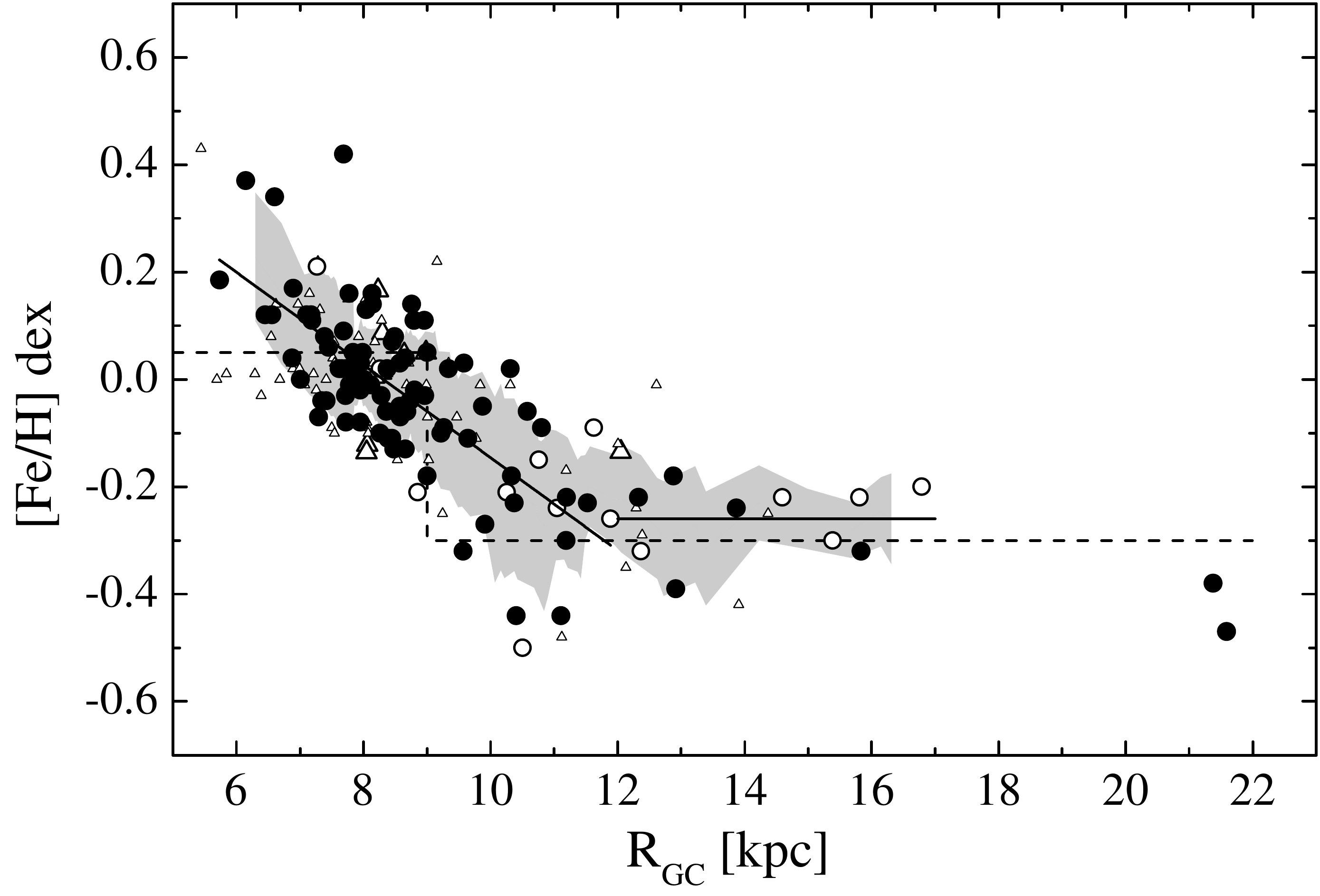}}
\caption{RMD of the open clusters. The circles represent spectroscopic data (HQS in black). Photometric metallicities are shown with triangles, with larger symbols for results that are based on at least two photometric systems. The dashed line shows the metallicity plateaus and the step-like discontinuity found by \citet{lepine11}, and the solid lines are the fits to our spectroscopic cluster data using the complete sample for the inner disc and the mean value for the outer area (see Table \ref{tab:gradients}). The grey area is the error range of the mean metallicity that we derived with a running average on the complete spectroscopic sample, as discussed in the text. }
\label{fig:rgcall}
\end{figure}

\citet{twarog97}, or more recently \citet{lepine11}, provided probably the most extreme interpretation of the RMD with two shallow plateaus and a step of $\sim$0.3\,dex at \rgc\ $\sim$ 9\,kpc\footnote{Based on \rgc$_{,\sun}$ = 8\,kpc.  \citet{lepine11} used 7.5\,kpc, and \citet{twarog97} 8.5\,kpc as the solar distance.} (see dashed line in Fig. \ref{fig:rgcall}). \citet{lepine11} explained this as a consequence of the corotation ring-shaped gap in the density of gas and that the metallicity evolved independently on both sides. 

The spectroscopic open cluster data can be considered as the most accurate sample, and their use is the logical first step to analyse the RMD. We make use of both the HQS and the LQS sample because the latter contributes significantly to the knowledge of the outer Galactic area (see Fig. \ref{fig:rgcall}). Except for the two objects discussed in Sect. \ref{sect:metcomp}, we adopted the higher quality data if results in both categories were available. There is no significant change if they are combined in a weighted manner, for example by a down-weight of two thirds for the LQS results. 

Figure \ref{fig:rgcall} shows that there are two groups of objects at \rgc\ $\sim$ 7\,kpc and 11\,kpc\footnote{The overabundant inner HQS clusters NGC~6583, NGC~6253, NGC~6791, and the outer underabundant HQS/LQS objects Trumpler~5, NGC~2243, and NGC~2266 (both groups ordered by \rgc).} that exhibit a strong over- or underabundant behaviour compared to the other clusters. Both groups deviate by at least 3\,$\sigma$ from the mean metallicity of the other objects at a comparable Galactocentric distance. However, the overabundant inner group might be evidence for a very steep metallicity gradient in the innermost disc, as proposed for example by \citet{andriev02} or \citet{pedic09} based on Cepheid data. On the other hand, the underabundant group represents about 20\,\% of the objects at that distance. This shows one of the main problems in the analysis of the RMD with current still limited data sets: which objects are representative for a particular Galactic distance? 

It is possible to provide arbitrarily justified arguments for the exclusion of almost each object. An old age or high positions above the Galactic plane, for example, might be indicators that objects are far from their birthplaces. Furthermore, \citet{caffau14} speculated about an extragalactic origin of Trumpler~5, similar to the most distant open clusters Berkeley~29 and Saurer~1 \citep{carraro9}. NGC~2266 has a high probability to belong to the thick disc \citep{reddy13}.
Assigning a weight to objects that might not represent the chemical characteristic of a particular position (e.g. owing to migration or eccentric orbits), however, could lead to a misleading conclusion as well. A detailed analysis of the cluster orbits was performed for only a limited number of objects. NGC~6791 is probably the most outstanding object  for its age and metal-richness (7\,Gyr, \FeH\ $\sim$ 0.4\,dex, at \rgc\ $\sim$ 7.7\,kpc). The dynamical study by \citet{jilkova13} resulted in a low probability that the cluster originates from the very inner disc (\rgc\ = 3--5\,kpc) and migrated outward to the present-day location. However, owing to the eccentric orbit, the cluster spans Galactocentric distances from about 5\,kpc to 9\,kpc.

The age, which is known for all objects with acceptable accuracy, could be a useful criterion to select a proper subsample that reflects the true chemical property of a particular Galactic position. But the unequal distribution of the cluster ages does not allow tracing the RMD in detail; almost all objects younger than 500\,Myr are located within \rgc\,$\sim$\,9\,kpc. Another difficulty in the correct interpretation of the RMD might arise if there are as yet unobserved 'key' objects. For example, without the availability of data for the outermost clusters, a flattening of the metallicity gradient cannot be concluded at all. We therefore have to work with this biased sample, but we tried to overcome the difficulties by at least including (complete sample) and excluding (cleaned sample) the strong over- and underabundant objects in the analysis. A comparable approach by including and excluding single objects was already used by \citet{twarog97} to see the effects on the derived gradients.   

We used the quantitatively derived transition radius by \citet{yong12} as a limit to determine a linear fit of the RMD for the inner disc as given in Table  \ref{tab:gradients}. It is quite evident that the inclusion of the six deviating points results in a somewhat steeper gradient. Obviously, the derived gradients could be artificial if the proposed step function (see Fig. \ref{fig:rgcall} and discussion before) is real. We therefore split the data set into the distance ranges \rgc\,$\le$\,9\,kpc and 9\,$<$\,\rgc\,$<$\,12 and derived linear fits as well. As suggested by \citet{corder01}, the comparison of the results for both subranges might be a critical test for the existence of this considerable break and discontinuity in the RMD. Both subranges show a comparable slope within the errors, regardless of whether the deviating points are included or excluded. This means that the results favour a roughly constant decline of the metallicity and not a sharp break. Although the inclusion of the deviating points results in larger errors for the little sampled outer range, the correlation coefficient of 0.36 gives a randomness probability of only 8\,\%. For the inner range, a gradient exists without any doubt, and the probability that the gradient comes from a random sample is one per mille and even an additional factor of 50 lower if the deviating overabundant objects are included. However, the large number of objects at the solar circle clearly dominates the analysis. More cluster data beyond \rgc\ $\sim$ 9\,kpc would therefore be helpful to improve the results in that distance range.  

\begin{table}
\caption{Derived metallicity gradients based on open cluster data.} 
\label{tab:gradients} 
\centering 
\begin{tabular}{l l l l l} 
\hline\hline 
Range & N & ZP & Slope &  Notes\\ 
\hline
$<$\,12\,kpc & 88 & +0.72 $\pm$ 0.08 & $-$0.086 $\pm$ 0.009 & all data \\
$<$\,12\,kpc & 82 & +0.54 $\pm$ 0.07 & $-$0.066 $\pm$ 0.007 & \tablefootmark{a} \\
$\le$\,9\,kpc & 64 & +0.71 $\pm$ 0.14 & $-$0.085 $\pm$ 0.017 & all data \\
$\le$\,9\,kpc & 61 & +0.51 $\pm$ 0.12 & $-$0.061 $\pm$ 0.015 & \tablefootmark{a} \\
9 -- 12\,kpc & 24 & +0.52 $\pm$ 0.39 & $-$0.068 $\pm$ 0.037 & all data \\
9 -- 12\,kpc & 21 & +0.41 $\pm$ 0.29 & $-$0.054 $\pm$ 0.028 & \tablefootmark{a} \\
$>$\,12\,kpc & 12 & $-$0.04 $\pm$ 0.12 & $-$0.016 $\pm$ 0.007 & all data \\
$>$\,12\,kpc & 10 & $-$0.26 $\pm$ 0.07 &  & \tablefootmark{b} \\
\hline 
$\le$\,0.5\,Gyr & 35 & +0.62 $\pm$ 0.12 & $-$0.079 $\pm$ 0.015 & all data \\
1 -- 2.5\,Gyr & 18 & +0.74 $\pm$ 0.11 & $-$0.082 $\pm$ 0.013 & all data \\
1 -- 2.5\,Gyr & 17 & +0.63 $\pm$ 0.12 & $-$0.072 $\pm$ 0.013 & \tablefootmark{a} \\

\hline
\end{tabular}
\tablefoot{Zero points (ZP) and slopes of the linear fits.\tablefoottext{a}{Excluding the strong over- or underabundant objects as discussed in the text.} \tablefoottext{b}{Mean metallicity excluding the most distant objects Berkeley~29 and Saurer~1.}}
\end{table}

Even fewer objects cover the outer disc (\rgc\ $\ge$ 12\,kpc), and this sample also includes the most distant objects Berkeley~29 and Saurer~1, which are well separated from the other clusters. We derived a mean metallicity excluding the two distant objects (see Table \ref{tab:gradients} and Fig. \ref{fig:rgcall}), but provide a linear fit over the complete distance range as well. The mean value intersects with the gradients derived for the inner disc at \rgc\,$\sim$\,11.3\,kpc or $\sim$\,12.2\,kpc, depending on whether the gradient for the complete or for the reduced sample is used. These values agree with other studies of the transition radius \citep[e.g.][]{Carraro07,Friel10,yong12}. 

\begin{figure}
\centering
\resizebox{\hsize}{!}{\includegraphics{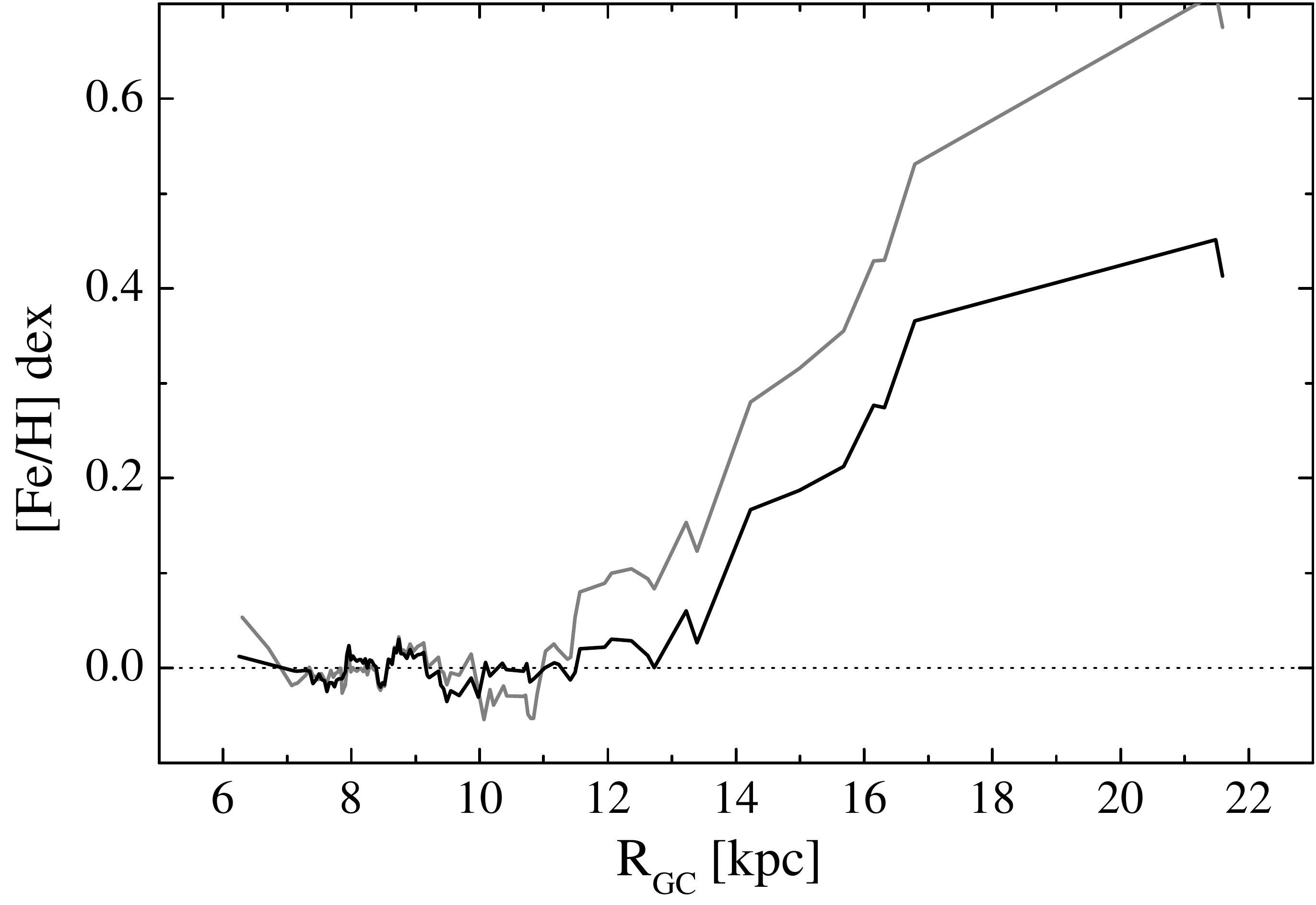}}
\caption{Residuals of the running average for the complete (black line) and cleaned sample (grey line) after subtracting the respective gradients of the inner disc objects. }
\label{fig:running}
\end{figure}

To further investigate the RMD, we employed a running average by starting with the innermost cluster. This approach was recently used by \citet{genovali14} to study the fine structure of the metallicity gradient based on Cepheids. We grouped the sample either by a constant number of 15 clusters or by a maximum distance range of 1\,kpc, whichever criterion was fulfilled first. This kind of selection takes the well-sampled area around the solar circle into account, but also less populated Galactocentric distances. The defined groups cover a mean distance range of 0.5\,kpc close to the solar circle (7--9\,kpc), but the number of included  objects per group drops to fewer than five beyond \rgc\,$\sim$\,11.5. We derived the running average and the standard deviations for the complete sample and for the list that excludes the deviating objects. Figure \ref{fig:running} shows the residuals after subtracting the respective gradients for the range \rgc\,$<$\,12\,kpc. The values estimated above for the transition radii are also confirmed by this figure. However, as already mentioned, the coverage of objects at this distances is poor, and the transition between inner and outer disc is most likely a smooth zone and not a strict position. Nevertheless, there are no significant deviations out to the transition radius (or zone), again suggesting a constant decline of the metallicity in the inner disc. 

The current available open cluster data do not allow a conclusion about an additional, probably steeper, metallicity gradient in the innermost disc, as it was concluded based on Cepheid data. If the additional gradient exists, it will probably start at \rgc\,$\lesssim$\,7\,kpc.

\subsection{Age-metallicity relation}
\label{sect:amr}
Another topic in Galactic research received much attention in recent years -- the age-metallicity relation (AMR). However, based on open cluster data, the conclusions remained in principle the same as was reached by \citet{carraro94}: the metal content of a cluster does not seem to depend on its age. This also holds for more recent studies that were based on high-resolution spectroscopic samples \citep[e.g.][]{Friel10,yong12}. \citet{pancino10} only noted a very mild decrease of metallicity with age. Thus, a strict AMR does not exist in the Galactic disc, but there is a tendency that older clusters have on average  lower metallicities \citep{magrini09}. Recently, \citet{bergem14} used high-resolution spectroscopic data of a sample of field stars and found a clear decline in metallicity for stars older than 8\,Gyr. 

\begin{figure}
\centering
\resizebox{\hsize}{!}{\includegraphics{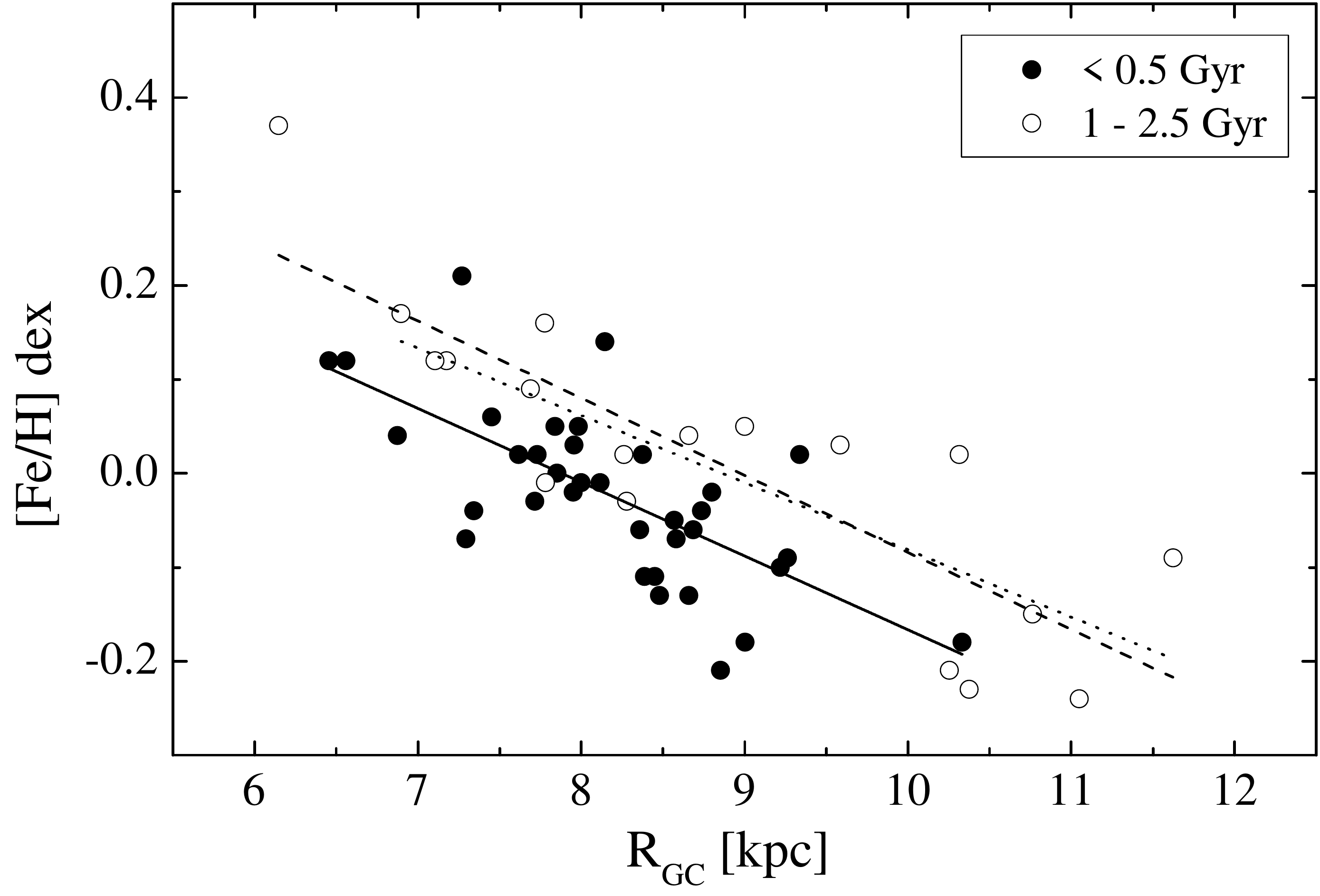}}
\caption{RMD of the open clusters, grouped by the age. The black solid line is the linear fit of the young group, the dashed and dotted lines are the results for the old group using the complete and cleaned sample, respectively.  The parameters are given in the lower panel of Table \ref{tab:gradients}. }
\label{fig:rgc_age}
\end{figure}

Our cluster sample unfortunately includes only a few objects that are close to that age. Although no striking evidence for the AMR was found at a younger age so far, we used our large sample to validate the previous results. The inner disc out to about 12\,kpc is the only area that includes objects of a broad age range, but most of the youngest clusters are found closer than \rgc\,$\sim$\,9\,kpc. Apart from metallicity, the age of open clusters is certainly the most inaccurate cluster parameter. The use of mean values most probably reduces the influence of single erroneous results for a cluster, but such an inhomogeneous compilation might include unknown systematic distortions and biases. Still, significant efforts have to be made to derive the age and all other cluster parameters on a homogeneous scale for numerous open clusters. All large homogeneous surveys that are available to date use solar metallicity isochrones
for simplicity. This work provides the basis to fix at least one parameter (the metallicity) in an isochrone fitting procedure, for example.  

However, our data certainly allow distinguishing between young and old objects and defining age groups. If an AMR exists, this feature will probably only be significantly  detected if the groups are well separated by the age and the covered age range in these groups  is sufficiently small to avoid an additional scatter in metallicity owing to the AMR. The use of groups that are well separated by age offers the additional benefit that the influence of erroneous results for the age is almost negligible. Furthermore, open clusters with a spectroscopically determined metallicity are probably among the best and most often studied objects of the whole known open cluster population.
 
We therefore used all objects out to 12\,kpc and defined two age groups ($<$\,0.5\,Gyr and 1--2.5\,Gyr) with a median age of 0.2 and 1.7\,Gyr. There are only 13 older objects, which span a broad age range from about three to eight Gyr, and most of the deviating objects belong to them. There is only one object in the defined older group (NGC~2360) that shows a large spread among the six compiled age determinations so that even a marginal interference with the young group might be possible. Although the most recent multicolour study of this object \citep{oralhan15} also pointed to an old age, we excluded this cluster from the group for the sake of reliability. The mean error of the age based on the compilation results in about 80 and 400\,Myr for the young and old group, respectively, and there is no overlap between the groups even if taking the full error ranges into account.  

We analysed the metallicity gradient for the two age groups with and without the deviating object in the old group (NGC~6583), as also done in the analysis of the RMD. The results are shown in Fig. \ref{fig:rgc_age} and listed in Table \ref{tab:gradients}. The derived gradients agree within the errors, but the metallicity level differs between the young and old group. At the solar circle, the metallicity increases from $-$0.01\,dex for the young group to +0.08\,dex or +0.05\,dex for the old group, using the complete and cleaned sample, respectively. The scatter of the cleaned samples ($\sigma$\,$\sim$\,0.07\,dex) is of the same order as the increase, but is somewhat lower than the value we obtain based on the whole cleaned sample without age grouping ($\sigma$\,$\sim$\,0.09\,dex). However, the errors of the linear fits in Table \ref{tab:gradients} do not allow us to assess the significance of the metallicity difference. 

It is evident that the gradient of the whole cleaned sample is much better defined because of the larger number of objects. After correcting for this gradient, we obtain mean residuals of $-0.024 \pm 0.069$\,dex (median $-$0.022\,dex) and $+0.044 \pm 0.077$\,dex (median +0.054\,dex) for the young and old group, respectively. To further reduce the influence of the gradient, we restricted the samples to the distance range \rgc\ = 7--9\,kpc. This results in similar mean residuals as for the whole distance range: $-0.025 \pm 0.072$\,dex (median $-$0.022\,dex, 27 objects) and $+0.041 \pm 0.054$\,dex (median +0.051\,dex, eight objects) for the young and old group, respectively.

The typical error of metallicity in our sample is 0.06\,dex, which is of the same order as the scatter and the apparent increase of metallicity. However, there is a mixture of HQS and LQS results, and the metallicity of several clusters is based on few or even on only a single star. When we repeat the last analysis, but using only HQS data that are based on more than two stars, we derive mean metallicities of $-0.025 \pm 0.041$\,dex (median $-$0.020\,dex, 14 objects) and $+0.044 \pm 0.058$\,dex (median 0.054\,dex, seven objects) for the young and old group, respectively. There is a further improvement of the scatter in both groups, but it was based on a small number of objects in the first place.   

To test if the residuals of the young and old clusters are statistically different or not, a Wilcoxon sign 
rank, a Kolmogorov-Smirnov, and a Student's t-test \citep{Hill05} was performed. The difference of the methods 
are that the Wilcoxon sign rank and Kolmogorov-Smirnov tests also account for a non-normal data distribution. 
We have to emphasize that we work here with small number statistics and that the tests do not take the errors into account. However, all three tests resulted in a difference of both data sets with a significance of at least 95\%. From a statistical point of
view and keeping the limitations in mind, we can conclude
that the data allow the interpretation as an increase of the metallicity level with age.

We cannot exclude selection effects owing to the age groups, for example. To further validate the results, a homogeneous age scale for the open clusters has to be derived. This will probably allow defining more and better defined age groups and tracing the evolution of this effect with age. Such an analysis is certainly limited by the current accuracy of the metallicity determinations.  Furthermore, the sample of older clusters at the solar circle has to be enlarged because their number is about half of the younger population. Finally, we note that an increase of metallicity with age can be explained by radial migration. Recent simulations \citep{Grand15,minchev13} confirmed the broadening of the metallicity distribution owing to the age, and radial outward migration appears to be the dominating effect.

\subsection{Usefulness of photometric metallicities}

Until now we have only considered spectroscopic metallicities, but these represent just about 60 percent of our compiled total sample. The use of metallicities based on a single photometric system might be too inaccurate, particularly if a cluster shows non-solar scaled element abundances (see also discussion in Sect. \ref{sect:metcomp}), and it might  therefore distort the results that we obtained with the spectroscopic metallicities. However, in addition to the adopted spectroscopic sample, there are only nine clusters with at least two individual photometric metallicity estimates. All of them but one (NGC~2141) are located very close to the solar circle (\rgc\, = 8 -- 9\,kpc), an area that is already well covered by spectroscopic data (see Fig. \ref{fig:rgcall}). We therefore refrain from a complete re-analysis that includes these few objects, but instead briefly discuss the benefit of using photometric metallicities. We note that even after evaluating the Washington data and recovering the photoelectric \ubv\ metallicities, the sample size will be increased by not more than seven clusters for which at least two estimates are available (all at \rgc\,$\sim$\,8\,kpc). 

\begin{figure}
\centering
\resizebox{\hsize}{!}{\includegraphics{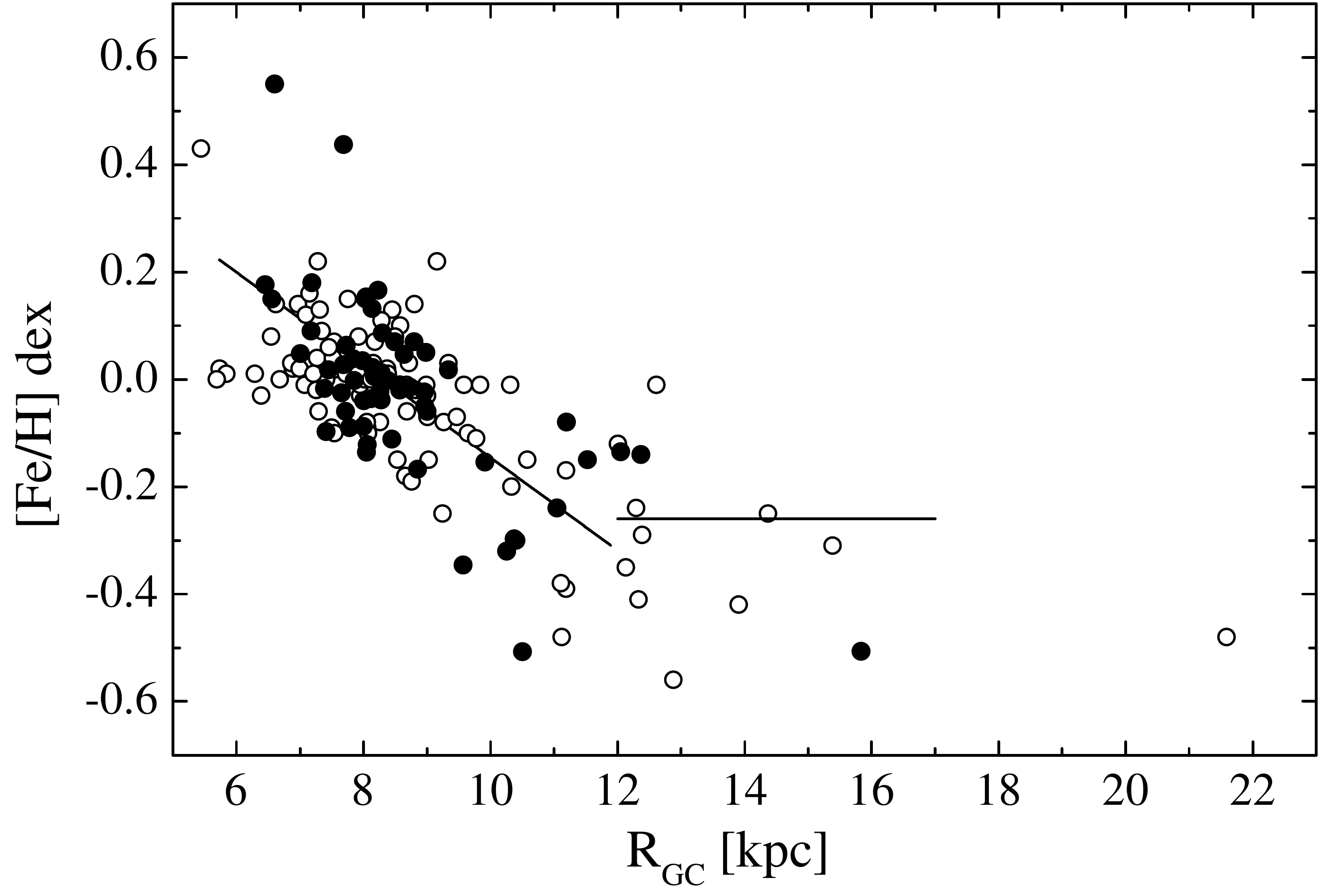}}
\caption{RMD of the open clusters exclusively based on the photometric results. Filled circles represent mean metallicities based on at least two individual estimates, and open circles show the remaining sample with only one photometric determination.  }
\label{fig:rgc_phot}
\end{figure}

Figure \ref{fig:rgc_phot} shows the RMD that is only based on photometric metallicities of 149 open clusters. Compared to Paper I (Fig. 5 therein) a significant improvement is visible that
is due to the careful data selection and their evaluation. In particular the exclusion of the UBV$_c$ estimates has resulted in a clearer and narrower metallicity distribution.

In Fig. \ref{fig:rgc_phot} we overlaid the fits to the spectroscopic data that were shown in Fig. \ref{fig:rgcall} to allow a direct comparison between the photometric and spectroscopic results. The gradient in the inner disc is detected in the photometric data as well. However, the restriction to the N\,$\geq$\,2 sample leads to a somewhat steeper slope and to errors that are almost twice as large as those derived for the spectroscopic data. Obviously, this is caused by the poorly populated range at 10--12\,kpc and the large scatter of the datapoints. Furthermore, compared with the spectroscopic RMD, the underabundant group appears not to
be separated anymore. The larger scatter continues in the outer disc, where the photometric metallicities are considerably lower than the spectroscopic values. Although only single estimates are available for most objects, problems of the recalibration in the underabundant regime cannot be excluded
either; almost all data originate from the Friel and Ca {\sc ii} samples. At \rgc\,$\lesssim$\,7\,kpc we also detect some objects with solar metallicity that are located well below the gradient; these originate from the DG sample. This means that several photometric systems might be affected by different types of systematic uncertainties, which could be compensated for only by using proper mean values that are based on several individual estimates.

Finally, we note that the photometric metallicities do not show the small increase in metallicity with age that we discussed in Sect. \ref{sect:amr}. It is evident that the accuracy of these data is in general much lower than in spectroscopic results.  Although the number of spectroscopic investigations increased enormously in recent years, we do not consider photometric metallicities as completely outdated, they can still provide valuable information for objects that are difficult to study spectroscopically even with modern instruments.

\subsection{Comparison with Galactic chemical models}
\label{sect:models}
In addition to open clusters, the Cepheids are useful probes of the Galactic disc. Iron abundances based on high-resolution spectroscopic data are available for numerous objects, and they are therefore a valuable additional source to investigate the Galactic iron content. Recently, \citet{genovali14} combined their own and literature results to a total sample of 450 Cepheids. They homogenized the metallicity scale and recognized offsets of up to about 0.3\,dex between different well-covered data sets, comparable to some open cluster results discussed in Paper II (e.g. Collinder~261). We therefore excluded 11 Cepheids \citep[the data adopted by][]{sziladi07} because the overlap with other data sets is too small to derive a possible offset \citep[see][]{genovali14}. The authors mentioned that objects close to the Galactic plane show a significantly lower spread in iron, in particular in the outer disc. We therefore restricted the sample to objects that are located closer than 500\,pc to the Galactic plane, which excludes only 25 additional objects.

It is essential to compare the metallicity scale of the open clusters and Cepheids. The most direct way is to use Cepheids that are members of open clusters. However, \citet{anderson13} listed only 23 confirmed Cepheid cluster stars. Four of the listed clusters are included in our sample: IC~4725, NGC~5662, NGC~6067, and NGC~6087, with two Cepheid members in NGC~6067. Spectroscopic (LQS) data are available for only one of these clusters (NGC 6087, hosting the Cepheid S Nor). Because the spectroscopic cluster metallicity is based on the same star, a comparison cannot be considered as independent. We therefore adopted the photometric value for NGC~6087. Based on the five matches, we calculated a mean difference (Cepheids $-$ clusters) of +0.05$\pm$0.08\,dex (median +0.06\,dex). Thus, the adopted metallicity scale by \citet{genovali14} is probably  slightly higher than the cluster scale. However, the small number of objects and the use of photometric data does not allow a final conclusion.

We therefore compared the mean metallicities of the open clusters and Cepheids close to the solar position (\rgc\,=\,7.5--8.5\,kpc). Only young open clusters ($<$\,200\,Myr) were considered because of the young age of Cepheids. We note that the Cepheids are all younger than about 150\,Myr in the period-age relation by \citet{bono05}. The mean difference between the two samples is +0.08$\pm$0.11\,dex, another hint of the higher metallicity scale of the Cepheids. 
We corrected the metallicities of the Cepheids by this difference, and they now roughly represent the metallicity scale by \citet{luck11} as shown in the comparison by \citet{genovali14}. Nevertheless, the probably more accurate way to match the scale is by deriving more metallicities of open clusters that host Cepheids, but based on photometric methods or preferably spectroscopic data of red giants or main-sequence stars. 

The metallicity gradient of the Cepheid sample was investigated
in detail by \citet{genovali14}, who derived slopes
between $-$0.051 and $-$0.060 dex kpc$^{-1}$, depending on the restrictions
and subsamples that were used (e.g. by excluding deviating results or
using stars close to the galactic plane). These values are compatible  with our results based on the cleaned open cluster sample, whereas the use of the complete data set always yield somewhat steeper slopes. However, the most apparent difference in the RMD of Cepheids and open clusters is that Cepheids show a continuous decline of metallicity out to \rgc\,$\sim$\,18\,kpc.

\begin{figure}
\centering
\resizebox{\hsize}{!}{\includegraphics{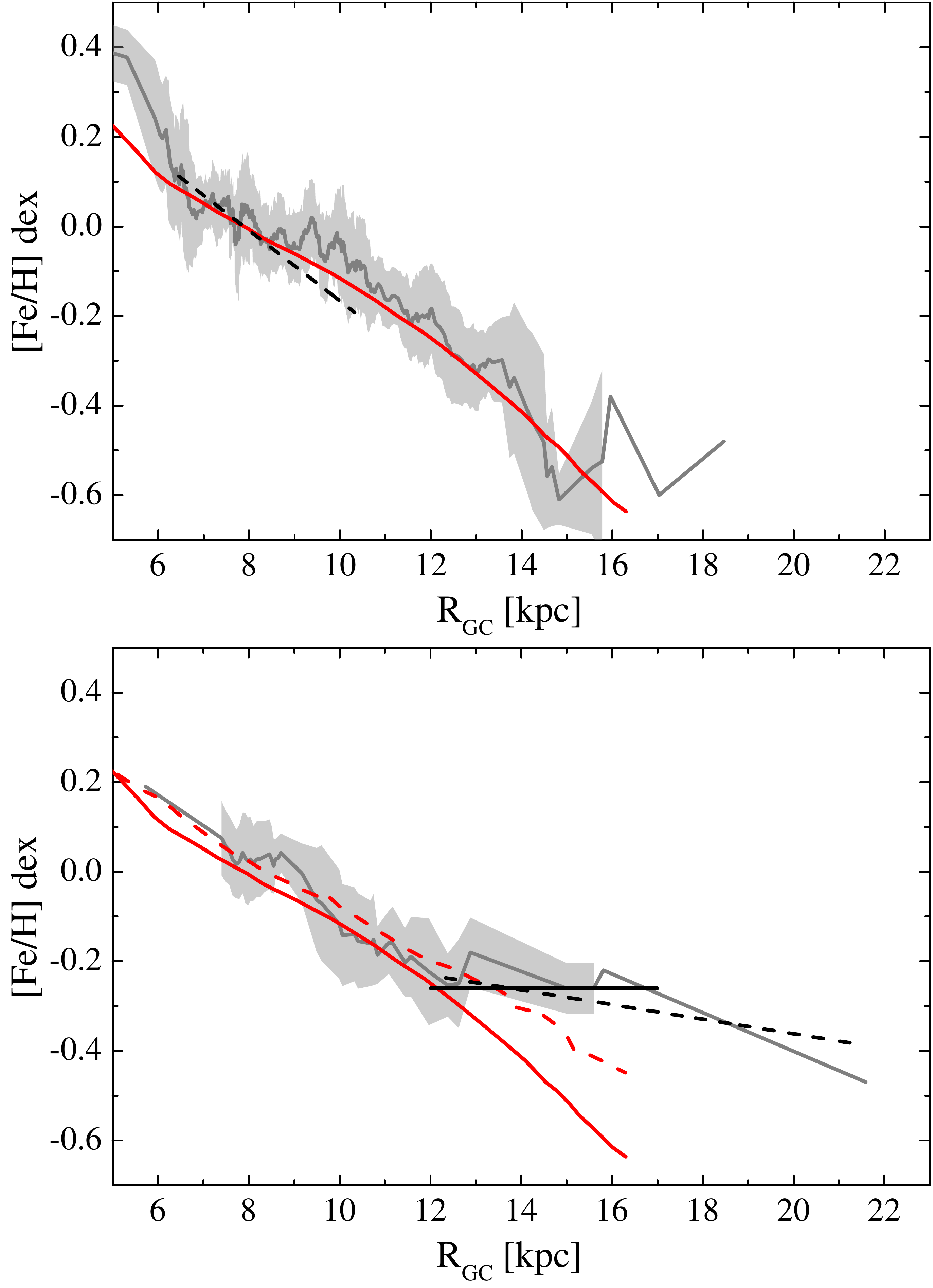}}
\caption{RMD of the open clusters and Cepheids compared to Galactic chemical models. Upper panel: The running average and the spread in iron of the Cepheid sample is shown as a grey line and a light grey area. The red line is the chemical model for the gas by \citet{minchev13}, the black dashed line shows the gradient derived for the youngest cluster age group (Table \ref{tab:gradients}). Lower panel: In grey we show the running average of intermediate-age clusters (0.5 -- 3\,Gyr), the red dashed line is the chemodynamical model by \citet{minchev13} for 2\,Gyr, the black solid line is the mean value of outer disc clusters, and the black dashed line is the gradient that we derived for all clusters in the outer disc.}
\label{fig:modelcomp}
\end{figure}

We used the Cepheid data to derive a running average with the same criteria that we adopted for the open clusters in Sect. \ref{sect:rmd}. We did not detect any significant difference by using a better sampling. The upper panel of Fig. \ref{fig:modelcomp} compares the results for open clusters with the Cepheids. Obviously, the gradient of the youngest cluster group is somewhat steeper than the one based on Cepheids. The slope of the cluster groups does not show a clear dependence on age, thus the steeper gradient might be attributed to the smaller sample and the narrower distance range that is covered by the young clusters. The sample of the Cepheids represents a very young population with a mean age of about 60\,Myr. Thus, they have not moved much from their birthplace, and a comparison with a Galactic chemical model for the gas is justified. Figure \ref{fig:modelcomp} shows that the model by \citet{minchev13} agrees very well with the observations. We shifted the model by about 0.05\,dex towards lower metallicities to match the Cepheids and young open clusters at the solar position. 

\citet{genovali14} identified some candidate
Cepheid groups that show an enhanced metallicity by about 0.05 dex (the groups VI to X in their paper) and are located close to the Perseus spiral arm. These can also be identified in Fig. \ref{fig:modelcomp} at \rgc\,$\sim$\,9--10\,kpc. The Cepheids also show an enhancement of the metallicity in the innermost Galactic region with a much steeper gradient than the chemical model predicts. These objects could be related to the Carina–Sagittarius arm. The spiral arms are the star-forming machines of the disc, thus the effect
of the spiral arms cannot be ignored \citep{lepine11}. However, a simple extrapolation of the metallicity towards the Galactic centre is not justified because inner disc and Galactic bulge and bar show different trends \citep[see discussion by][and references therein]{genovali13}.

\citet{minchev13} also presented a model that includes the dynamical effects of the Galactic disc on the RMD after 2\,Gyr. This makes it appropriate for a comparison with older open clusters, which are shown in the lower panel of Fig. \ref{fig:modelcomp}. For the sake of consistency, the chemodynamical model was shifted by the same value to lower metallicities as discussed above for the initial chemical model. Figure \ref{fig:modelcomp} also shows a running average of clusters in the age range 0.5--3\,Gyr (44 objects of the cleaned sample), but by grouping only ten objects. Unfortunately, the number of available clusters does neither allow a better sampling, nor does it define a narrower age range. First of all, the comparison of the two models supports our findings in Sect. \ref{sect:amr}, namely the increase of metallicity with age owing to radial migration, but the increase indicated by the models (about 0.03\,dex at the solar circle) is only about half of the observed value. Furthermore, the gradient out to about 12\,kpc is little affected in the considered time interval of 2\,Gyr \citep[see discussion by][]{minchev13}. The metallicity levels of the models diverge at larger distances, which can also
be observed in a comparison of Cepheids as young population with open clusters as a much older one. However, the open clusters show a somewhat higher metallicity and a flatter gradient than the chemodynamical model predicts. This means that radial migration might have a stronger influence than adopted in the model or that there is another mechanism acting that was not considered as yet.

The knowledge of metallicity for more outer disc open clusters
will be important to further verify and to improve
chemodynamical models for the Galaxy. The updated open cluster
catalogue by \citet[][version 3.4]{dias02} lists about 40 objects that
are more distant than \rgc\,$\sim$\,13 kpc, of which about one fourth is included in our list of spectroscopic metallicities. There are two cluster candidates
(FSR~338 and Teutsch~12) that could be of special interest because of
their distances (\rgc\,$\sim$\,17 kpc) and the comparatively young age
of about 130 and 300 Myr, respectively.

In Paper II we tested several other Galactic models with the basic cluster sample (most of the HQS clusters shown here), but none of the models succeeded in predicting the observed metallicity gradient. Only one of the models \citep{schonr09} includes the effect of radial migration. The differences to the model adopted here are discussed in detail by \citet{minchev13}. These are among others that \citet{schonr09} have not considered the presence of a central bar and assumed that the vertical velocity dispersion of migrating populations is constant. 

\subsection{Azimuthal gradient}

\begin{figure}
\centering
\includegraphics[width=90mm]{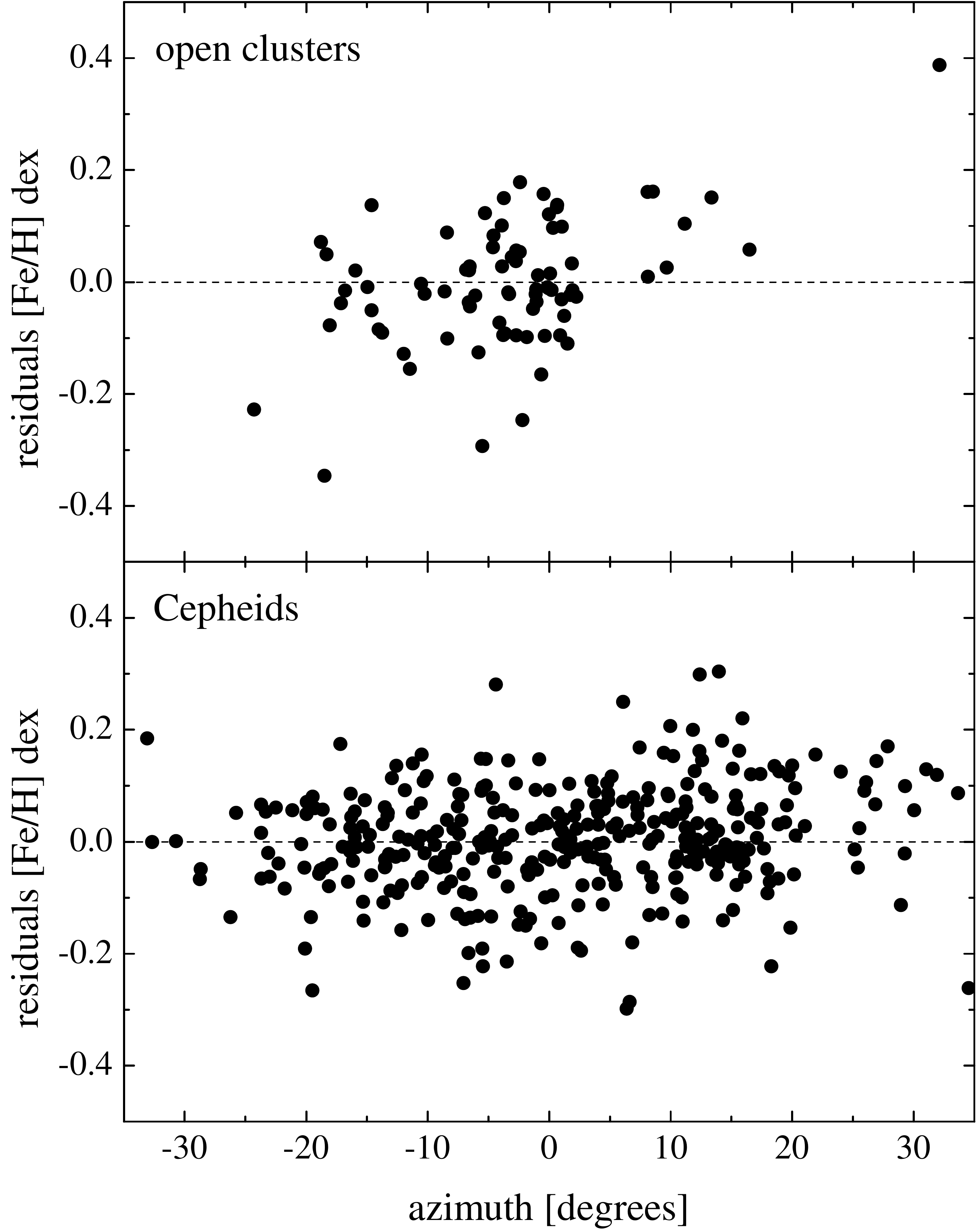}
\caption{Residuals of the iron abundance of open clusters and Cepheids with \rgc\ $\geq$ 7\,kpc as a function of the Galactocentric angle in the Galactic plane (azimuthal angle). Positive and negative azimuthal angles generally correspond to the second and third Galactic quadrants, respectively.  }
\label{fig:azimuth}
\end{figure}

We used the cluster and Cepheid data to investigate whether the metallicity shows a dependency on the Galactic position in addition
to the dependence on the radial gradient. This could be the position in the Galactic quadrants, for example. Figure \ref{fig:azimuth} shows the residuals of the iron abundance of Cepheids and open clusters as a function of the Galactocentric angle in the Galactic plane (azimuthal angle). From the cluster data we subtracted the gradient that was derived for the cleaned sample, and for the Cepheid data we adopted the gradient that \citet{genovali14} derived for the sample that is located closer to the Galactic plane. The inner distance limit was set to \rgc\ = 7\,kpc to exclude the Galactic area that probably shows the onset of a steeper slope (see upper panel of Fig. \ref{fig:modelcomp}). Thus, the data cover most of the second and third Galactic quadrants. For the open clusters we used the outer limit of the gradient at 12\,kpc, whereas we also included more distant Cepheids because the derived gradient extends over the complete covered distance range \citep[see][]{genovali14}.

The total cluster sample of 172 objects shows that twice as many open clusters are located in the third galactic quadrant than in the second. The spectroscopic sample shows an even larger difference in the distribution. In particular for the farther clusters this could be a result of the interstellar absorption window at Galactic longitudes from 215\degr\ to 255\degr\ \citep[][]{vazquez08}. 

Figure \ref{fig:azimuth} shows that open clusters tend to have a higher metallicity in the second Galactic quadrant. However, the small number of objects in this area and age effects owing to radial migration do not allow a conclusion. We note that the objects deviating most strongly are once again NGC~2243, NGC~2266, NGC~6791, and Trumpler~5 (see Sect. \ref{sect:rmd}). The Cepheids, which are more numerous and much younger, can probably provide more details on differences between the quadrants. At larger positive angles (\ga 25\degr) the data give the impression of a higher abundance level. We inspected the data in 1\,kpc wide concentric rings by using both the residuals and the original metallicity. We found that the overabundant appearance is in principle caused by about ten objects that are somewhat closely spaced in the middle between the Perseus and the Carina-Sagittarius arms at a Galactic longitude of $\sim$\,70\degr\ and \rgc\,$\sim$\,7.5\,kpc \citep[or at X/Y $\sim$ 3.5/$-$1.5\,kpc in Fig. 5 by][]{genovali14}. Thus, they might belong to the Local arm (Orion spur). The other overabundant Cepheid groups by \citet{genovali14} can also be identified in Fig. \ref{fig:azimuth} at an azimuthal angle of about 15 degrees. We did not find any other evidence for an azimuthal dependency, which confirms the results of previous studies  \citep[e.g.][]{luck11,genovali14}.

\begin{figure}
\centering
\includegraphics[width=90mm]{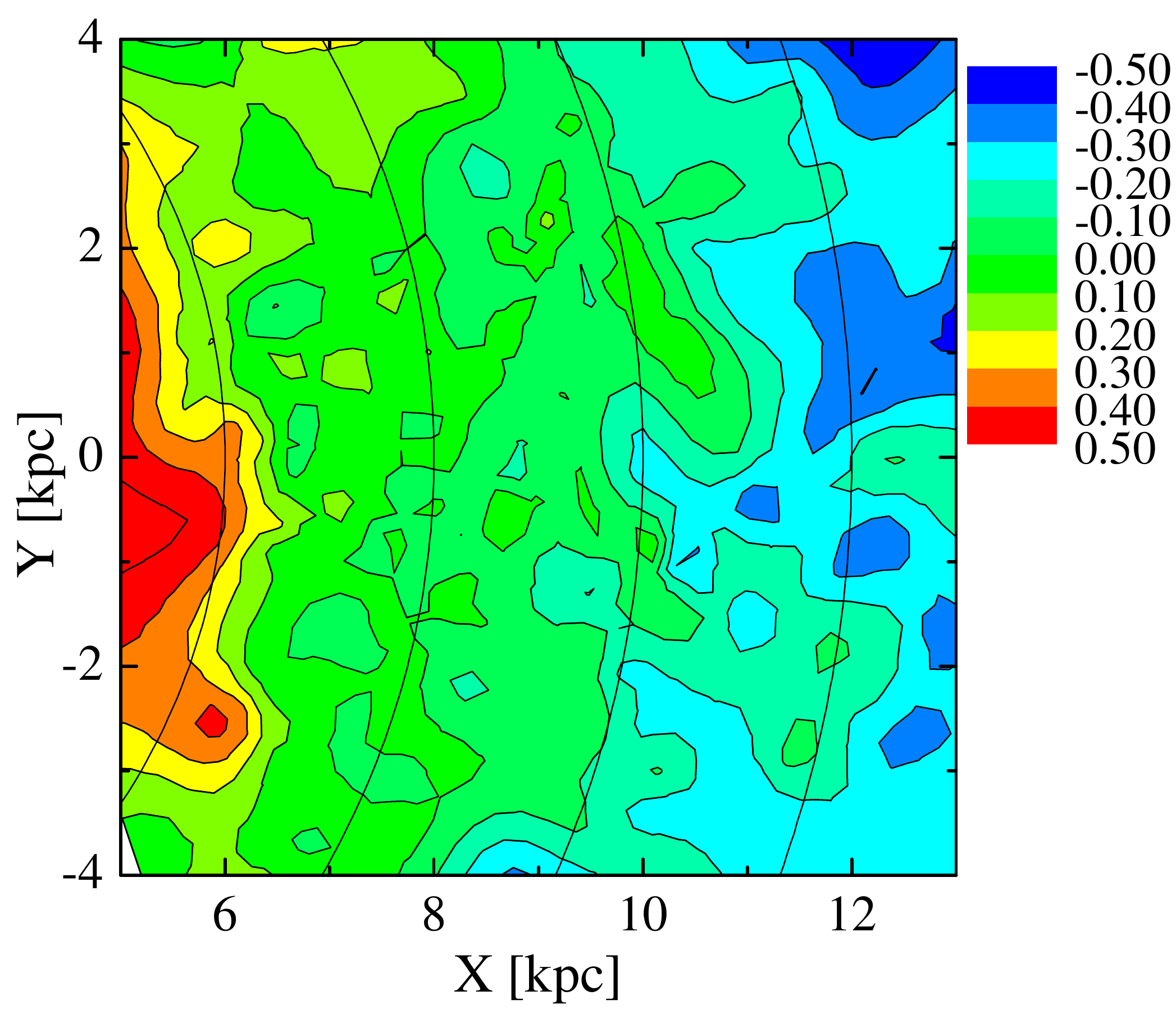}
\caption{Contour plot of the metallicity distribution in the X/Y plane using a combined sample of Cepheids and open clusters. The orientation is as in Fig. \ref{fig:xydist}. Solid lines indicate distances from the Galactic centre of 6, 8, 10, 
and 12 kpc.}
\label{fig:contour}
\end{figure}

Finally, we present a contour plot of the metallicity distribution in the Galactic plane by combining both spectroscopic data samples (see Fig. \ref{fig:contour}). This 8\,kpc$^2$ area is well covered by almost 500 objects. The contour plot shows the smooth decrease of metallicity towards the Galactic anticentre and that no other obvious features are present. This confirms the results found by \citet[][Fig. 5 therein]{luck11}, which were based on a smaller sample (about half the one used here). The contour plot based on photometric metallicities that we presented in Paper I (Fig. 4 therein) shows a much more patchy structure in a 4\,kpc$^2$ area around the Sun and a broader metallicity range as well. We conclude that this appearance was caused in the previous paper by the $\ubv_c$ metallicities that appear much more underabundant and inaccurate compared to all other methods and references.

\section{Summary and conclusions}
\label{sect:conclusion}
In the third paper of our series, which deals with the current knowledge of open cluster metallicities, we presented metallicities of 172 open clusters. We combined spectroscopic determinations with photometric estimates that were brought to a common scale. Most photometric methods are reasonably well scaled at this point, but some show offsets as large as about 0.2\,dex. Although for the sake of homogeneity and accuracy we had to exclude some available data sets, this compilation is currently by far the largest available sample of homogenised open cluster metallicities. These data provide a basis for several other studies, for example, statistical treatments of the Galactic cluster population, or  evolutionary studies of individual star groups in open clusters. The four cluster parameters (distance, age, reddening, and metallicity) are strongly coupled in the isochrone fitting procedure. Thus, our data set also allows fixing at least one parameter (the metallicity) and improving the accuracy of the others. All large open cluster parameter compilations that were derived in a homogeneous way still adopt solar metallicity isochrones \citep{netopil15}.

We used the spectroscopic metallicities of 100 open clusters for a detailed study of the radial metallicity distribution and derived a gradient of about $-$0.07\,dex\,kpc$^{-1}$ out to \rgc\,$\sim$\,12\,kpc, in agreement with several studies. Both the use of distance subsamples and of a running average confirm the continuous decline of metallicity. The use of age groups in the inner disc revealed an increase of the metallicity level with age. The older clusters ($\sim$\,1.7\,Gyr) appear more metal rich by about 0.07\,dex than clusters of a young group ($<$\,0.5\,Gyr). This is clearly the opposite of what was commonly looked for in the past, a decline of metallicity with age, but it might be an observational confirmation of radial migration. However, to further validate this result, a homogeneous age scale for the clusters and a somewhat larger sample of older objects are needed.

Furthermore, we compared open clusters and Cepheids with recent Galactic chemical and chemodynamical models. The models reproduce the observed increase of metallicity with age and generally agree
well with the observations of young and older populations. However, the comparison shows that the models probably underestimate the effect of radial migration, or that there is another not yet included mechanism acting in the outer disc, where open clusters show a much flatter gradient than predicted by the models. However, to study the reliability of the current models in more detail and to further improve them, two questions apparently need to be resolved
first: a somewhat larger open cluster sample that covers a broader Galactic distance range in each age group is needed. Most studied objects are still concentrated in the solar circle. Second, a more detailed investigation of the relative metallicity scale between open clusters and Cepheids is required. This should include an improvement of the knowledge of the metallicities of Cepheids that are cluster members and of the metallicities of the hosting open clusters, but the latter based on photometric methods or spectroscopic data of red giants or main-sequence stars.

Finally, we showed the Galactic metallicity distribution of a 8\,kpc$^2$ large area in a contour plot and the metallicity as a function of the azimuthal angle. Both confirm that there is little difference between the Galactic quadrants. However, there are some slightly overabundant areas that might be related to the spiral structure of the Galaxy.

The final results of spectroscopic surveys, such as the \textit{Gaia}-ESO Public Spectroscopic Survey or the \textit{Gaia} mission itself, will certainly provide the basis for more detailed studies. Furthermore, the availability of current and future homogeneous photometric datasets will allow complementing the spectroscopic metallicities with results obtained by photometric methods. We wish to mention in particular the DG method, which can be applied to any photometric data with available temperature calibrations.

\onllongtab{
\begin{landscape}
\begin{longtable}{lcclcclccclccl}
\caption{\label{tab:longtable} Metallicities of the open clusters}\\
\hline\hline
Cluster & [Fe/H] HQS & $\sigma$HQS & N &[Fe/H] LQS & $\sigma$LQS & N & [Fe/H] Phot & $\sigma$Phot & W & Systems & \rgc & Age & N \\
       & dex & dex &   & dex & dex &   & dex & dex &   &   & kpc & Gyr &  \\
\hline
\endfirsthead
\caption{Continued.}\\
\hline\hline
Cluster & HQS & $\sigma$HQS & N & LQS & $\sigma$LQS & N & Phot & $\sigma$Phot & W & Systems & \rgc & Age & N \\
       & dex & dex &   & dex & dex &   & dex & dex &   &   & kpc & Gyr &  \\
\hline
\endhead
\hline
\endfoot
\hline
\endlastfoot
Alessi 13 &   &   &   &   &   &   & +0.15 &   & 1.5 & 1 & 8.03 $\pm$ 0.00 & 0.53 $\pm$ 0.00 & 2 \\
Berkeley 11 &   &   &   &   &   &   & $-$0.01 &   & 1.5 & 1 & 10.31 $\pm$ 0.32 & 0.08 $\pm$ 0.06 & 7 \\
Berkeley 17 & $-$0.06 &   & 1 &   &   &   & $-$0.15 &   & 1 & 3 & 10.58 $\pm$ 0.44 & 7.79 $\pm$ 2.90 & 7 \\
Berkeley 18 & $-$0.39 &   & 1 &   &   &   &   &   &   &   & 12.92 $\pm$ 0.71 & 4.03 $\pm$ 0.63 & 6 \\
Berkeley 20 & $-$0.32 & 0.12 & 2 &   &   &   & $-$0.51 & 0.09 & 1.5 & 3,4 & 15.84 $\pm$ 0.26 & 6.24 $\pm$ 1.93 & 6 \\
Berkeley 21 & $-$0.18 &   & 1 & $-$0.54 & 0.20 & 3 & $-$0.56 &   & 0.5 & 3 & 12.88 $\pm$ 1.20 & 1.42 $\pm$ 1.13 & 6 \\
Berkeley 22 & $-$0.24 &   & 1 & $-$0.32 & 0.19 & 2 &   &   &   &   & 13.87 $\pm$ 0.80 & 3.07 $\pm$ 1.08 & 7 \\
Berkeley 23 &   &   &   &   &   &   & $-$0.42 &   & 1 & 4 & 13.91 $\pm$ 0.86 & 1.13 $\pm$ 0.40 & 5 \\
Berkeley 25 &   &   &   & $-$0.20 & 0.05 & 4 &   &   &   &   & 16.79 $\pm$ 1.78 & 5.08 $\pm$ 0.95 & 4 \\
Berkeley 26 &   &   &   &   &   &   & $-$0.35 &   & 0.5 & 4 & 12.13 $\pm$ 2.41 & 1.77 $\pm$ 1.62 & 4 \\
Berkeley 29 & $-$0.47 & 0.02 & 2 & $-$0.32 & 0.04 & 3 & $-$0.48 &   & 1.5 & 1 & 21.59 $\pm$ 3.14 & 2.67 $\pm$ 1.48 & 10 \\
Berkeley 31 &   &   &   &   &   &   & $-$0.25 &   & 1 & 3 & 14.37 $\pm$ 1.75 & 3.78 $\pm$ 2.63 & 8 \\
Berkeley 32 & $-$0.30 & 0.05 & 11 &   &   &   & $-$0.39 &   & 1 & 3 & 11.19 $\pm$ 0.52 & 3.85 $\pm$ 0.84 & 12 \\
Berkeley 33 &   &   &   & $-$0.26 & 0.05 & 5 &   &   &   &   & 11.89 $\pm$ 0.81 & 0.84 $\pm$ 0.27 & 7 \\
Berkeley 39 & $-$0.23 & 0.05 & 7 & $-$0.20 & 0.03 & 18 & $-$0.15 & 0.14 & 2 & 3,4 & 11.53 $\pm$ 0.40 & 6.77 $\pm$ 1.91 & 7 \\
Berkeley 70 &   &   &   &   &   &   & $-$0.01 &   & 1 & 4 & 12.62 $\pm$ 0.41 & 3.05 $\pm$ 0.73 & 4 \\
Berkeley 73 &   &   &   & $-$0.22 & 0.10 & 2 &   &   &   &   & 15.81 $\pm$ 1.24 & 1.71 $\pm$ 0.35 & 5 \\
Berkeley 75 &   &   &   & $-$0.22 &   & 1 &   &   &   &   & 14.59 $\pm$ 1.37 & 2.79 $\pm$ 0.98 & 4 \\
Berkeley 81 & +0.19 & 0.06 & 2 & +0.24 & 0.05 & 10 & +0.02 &   & 0.5 & 4 & 5.73 $\pm$ 0.27 & 0.95 $\pm$ 0.09 & 9 \\
Berkeley 99 &   &   &   &   &   &   & $-$0.48 &   & 1 & 4 & 11.12 $\pm$ 0.26 & 2.90 $\pm$ 0.42 & 6 \\
Blanco 1 & +0.03 & 0.07 & 6 &   &   &   &   &   &   &   & 7.96 $\pm$ 0.00 & 0.07 $\pm$ 0.06 & 8 \\
Collinder 110 & +0.03 & 0.02 & 3 &   &   &   & $-$0.01 &   & 1 & 4 & 9.59 $\pm$ 0.69 & 1.24 $\pm$ 0.62 & 5 \\
Collinder 140 &   &   &   &   &   &   & +0.01 & 0.04 & 2 & 1,5 & 8.17 $\pm$ 0.01 & 0.04 $\pm$ 0.01 & 9 \\
Collinder 173 &   &   &   &   &   &   & $-$0.10 &   & 1 & 5 & 8.07 $\pm$ 0.00 & 0.02 $\pm$ 0.01 & 2 \\
Collinder 258 &   &   &   &   &   &   & $-$0.09 &   & 0.5 & 2 & 7.50 $\pm$ 0.07 & 0.12 $\pm$ 0.08 & 4 \\
Collinder 261 & +0.00 & 0.04 & 13 &   &   &   & +0.05 & 0.05 & 2.5 & 1,3 & 7.01 $\pm$ 0.14 & 7.18 $\pm$ 2.63 & 6 \\
Collinder 272 &   &   &   &   &   &   & +0.03 &   & 1.5 & 1 & 6.87 $\pm$ 0.08 & 0.01 $\pm$ 0.01 & 10 \\
IC 166 &   &   &   &   &   &   & $-$0.17 &   & 0.5 & 3 & 11.19 $\pm$ 0.73 & 1.13 $\pm$ 0.61 & 7 \\
IC 1311 &   &   &   &   &   &   & $-$0.15 &   & 1 & 4 & 8.54 $\pm$ 0.41 & 1.38 $\pm$ 0.45 & 8 \\
IC 2391 & $-$0.01 & 0.03 & 12 &   &   &   & $-$0.09 & 0.05 & 2.5 & 1,5 & 8.00 $\pm$ 0.00 & 0.05 $\pm$ 0.03 & 11 \\
IC 2488 &   &   &   &   &   &   & +0.08 &   & 0.5 & 2 & 7.93 $\pm$ 0.00 & 0.19 $\pm$ 0.10 & 6 \\
IC 2602 & $-$0.02 & 0.02 & 7 & $-$0.05 & 0.04 & 6 & $-$0.03 &   & 1.5 & 1 & 7.95 $\pm$ 0.00 & 0.03 $\pm$ 0.02 & 12 \\
IC 2714 & +0.02 & 0.06 & 4 &   &   &   & $-$0.01 &   & 1 & 2 & 7.62 $\pm$ 0.01 & 0.28 $\pm$ 0.09 & 8 \\
IC 4651 & +0.12 & 0.04 & 18 &   &   &   & 0.09 & 0.01 & 2 & 2,5 & 7.17 $\pm$ 0.11 & 1.86 $\pm$ 0.73 & 17 \\
IC 4665 & $-$0.03 & 0.04 & 18 &   &   &   & $-$0.06 & 0.10 & 2.5 & 1,5 & 7.72 $\pm$ 0.01 & 0.04 $\pm$ 0.01 & 14 \\
IC 4725 &   &   &   &   &   &   & +0.00 &   & 1.5 & 1 & 7.41 $\pm$ 0.06 & 0.06 $\pm$ 0.03 & 8 \\
IC 4756 & +0.02 & 0.04 & 15 & $-$0.13 & 0.07 & 6 & $-$0.03 & 0.02 & 2 & 2,3 & 7.66 $\pm$ 0.04 & 0.66 $\pm$ 0.12 & 7 \\
King 1 &   &   &   &   &   &   & $-$0.01 &   & 1 & 4 & 8.99 $\pm$ 0.20 & 2.91 $\pm$ 1.49 & 4 \\
King 2 &   &   &   &   &   &   & $-$0.29 &   & 1 & 4 & 12.39 $\pm$ 0.48 & 4.92 $\pm$ 1.86 & 5 \\
King 5 &   &   &   &   &   &   & $-$0.11 &   & 1 & 3 & 9.78 $\pm$ 0.14 & 1.01 $\pm$ 0.21 & 6 \\
King 8 &   &   &   &   &   &   & $-$0.24 &   & 0.5 & 3 & 12.30 $\pm$ 1.20 & 0.62 $\pm$ 0.30 & 6 \\
King 11 &   &   &   &   &   &   & $-$0.07 &   & 1 & 3 & 9.47 $\pm$ 0.35 & 3.59 $\pm$ 1.65 & 8 \\
King 21 &   &   &   &   &   &   & +0.03 &   & 1.5 & 1 & 9.34 $\pm$ 0.37 & 0.02 $\pm$ 0.02 & 7 \\
Loden 807 &   &   &   &   &   &   & $-$0.10 &   & 0.5 & 2 & 7.54 $\pm$ 0.08 & 0.20 $\pm$ 0.00 & 2 \\
Lynga 1 &   &   &   &   &   &   & +0.02 &   & 1.5 & 1 & 6.89 $\pm$ 0.17 & 0.14 $\pm$ 0.05 & 6 \\
Melotte 20 & +0.14 & 0.11 & 2 &   &   &   & +0.02 & 0.01 & 2.5 & 1,5 & 8.14 $\pm$ 0.01 & 0.05 $\pm$ 0.03 & 13 \\
Melotte 22 & $-$0.01 & 0.05 & 10 &   &   &   & $-$0.04 & 0.11 & 2.5 & 1,5 & 8.12 $\pm$ 0.01 & 0.10 $\pm$ 0.04 & 13 \\
Melotte 25 & +0.13 & 0.05 & 61 & +0.14 & 0.04 & 21 & +0.15 & 0.03 & 3 & 1,2,5 & 8.04 $\pm$ 0.00 & 0.72 $\pm$ 0.13 & 7 \\
Melotte 66 & $-$0.32 & 0.03 & 6 & $-$0.28 & 0.03 & 7 & $-$0.35 & 0.03 & 4 & 1,2,3,4 & 9.57 $\pm$ 0.44 & 3.93 $\pm$ 1.17 & 8 \\
Melotte 71 & $-$0.27 &   & 1 &   &   &   & $-$0.15 & 0.02 & 3.5 & 1,5,6 & 9.91 $\pm$ 0.30 & 0.69 $\pm$ 0.40 & 10 \\
Melotte 105 &   &   &   &   &   &   & +0.06 &   & 1.5 & 1 & 7.45 $\pm$ 0.03 & 0.29 $\pm$ 0.10 & 17 \\
Melotte 111 & +0.00 & 0.08 & 10 &   &   &   & $-$0.04 & 0.03 & 2.5 & 1,5 & 8.01 $\pm$ 0.00 & 0.57 $\pm$ 0.10 & 10 \\
NGC 188 & +0.11 & 0.04 & 4 & $-$0.02 & 0.09 & 26 & $-$0.02 & 0.18 & 3 & 2,3,4 & 8.96 $\pm$ 0.12 & 6.27 $\pm$ 2.30 & 15 \\
NGC 559 &   &   &   &   &   &   & $-$0.25 &   & 1 & 4 & 9.25 $\pm$ 0.40 & 0.46 $\pm$ 0.16 & 8 \\
NGC 752 & $-$0.03 & 0.06 & 23 & $-$0.09 & 0.13 & 36 & $-$0.04 & 0.08 & 5.5 & 1,2,3,5,6 & 8.28 $\pm$ 0.03 & 1.69 $\pm$ 0.66 & 13 \\
NGC 1039 & +0.02 & 0.06 & 7 &   &   &   & +0.02 &   & 1.5 & 1 & 8.38 $\pm$ 0.02 & 0.19 $\pm$ 0.12 & 11 \\
NGC 1193 & $-$0.22 &   & 1 &   &   &   & $-$0.41 &   & 0.5 & 3 & 12.33 $\pm$ 0.76 & 4.77 $\pm$ 2.49 & 7 \\
NGC 1245 & +0.02 & 0.03 & 3 & $-$0.05 & 0.06 & 11 &   &   &   &   & 10.31 $\pm$ 0.26 & 1.07 $\pm$ 0.23 & 14 \\
NGC 1342 & $-$0.13 & 0.03 & 3 &   &   &   &   &   &   &   & 8.48 $\pm$ 0.12 & 0.42 $\pm$ 0.13 & 9 \\
NGC 1545 & $-$0.06 &   & 1 &   &   &   & $-$0.06 &   & 0.5 & 2 & 8.69 $\pm$ 0.07 & 0.28 $\pm$ 0.21 & 5 \\
NGC 1662 & $-$0.11 & 0.01 & 2 &   &   &   & $-$0.00 & 0.08 & 2 & 1,2 & 8.39 $\pm$ 0.03 & 0.38 $\pm$ 0.13 & 8 \\
NGC 1798 &   &   &   &   &   &   & $-$0.12 &   & 1 & 4 & 12.01 $\pm$ 0.54 & 1.48 $\pm$ 0.26 & 7 \\
NGC 1817 & $-$0.11 & 0.03 & 4 & $-$0.16 & 0.03 & 28 & $-$0.10 &   & 0.5 & 3 & 9.64 $\pm$ 0.21 & 0.82 $\pm$ 0.31 & 10 \\
NGC 1901 & $-$0.08 &   & 1 &   &   &   & $-$0.01 &   & 1.5 & 1 & 7.95 $\pm$ 0.01 & 0.77 $\pm$ 0.10 & 6 \\
NGC 1912 & $-$0.10 & 0.14 & 2 &   &   &   &   &   &   &   & 9.22 $\pm$ 0.12 & 0.26 $\pm$ 0.08 & 12 \\
NGC 1977 & $-$0.06 & 0.19 & 2 &   &   &   &   &   &   &   & 8.36 $\pm$ 0.08 & 0.01 $\pm$ 0.01 & 2 \\
NGC 2099 & +0.02 & 0.05 & 3 &   &   &   & +0.02 & 0.08 & 2.5 & 1,2 & 9.34 $\pm$ 0.06 & 0.36 $\pm$ 0.19 & 11 \\
NGC 2112 & +0.14 & 0.05 & 3 & +0.02 &   & 1 & $-$0.19 &   & 0.5 & 3 & 8.76 $\pm$ 0.08 & 3.84 $\pm$ 2.72 & 8 \\
NGC 2141 &   &   &   &   &   &   & $-$0.14 & 0.02 & 2 & 3,4 & 12.05 $\pm$ 0.15 & 2.19 $\pm$ 0.82 & 9 \\
NGC 2158 &   &   &   & $-$0.32 & 0.08 & 11 & $-$0.14 & 0.14 & 2 & 2,3 & 12.37 $\pm$ 0.58 & 1.67 $\pm$ 0.45 & 9 \\
NGC 2168 &   &   &   & $-$0.21 & 0.10 & 9 & $-$0.17 & 0.01 & 2 & 1,2 & 8.85 $\pm$ 0.12 & 0.10 $\pm$ 0.04 & 16 \\
NGC 2194 & $-$0.09 & 0.00 & 2 & +0.08 & 0.08 & 6 &   &   &   &   & 10.80 $\pm$ 0.43 & 0.60 $\pm$ 0.25 & 6 \\
NGC 2204 &   &   &   & $-$0.24 & 0.08 & 11 & $-$0.24 & 0.14 & 2 & 2,3 & 11.05 $\pm$ 0.45 & 1.85 $\pm$ 0.52 & 10 \\
NGC 2232 &   &   &   &   &   &   & +0.11 &   & 1.5 & 1 & 8.29 $\pm$ 0.02 & 0.04 $\pm$ 0.02 & 9 \\
NGC 2243 &   &   &   & $-$0.50 & 0.08 & 21 & $-$0.51 & 0.11 & 4 & 2,3,5,6 & 10.51 $\pm$ 0.22 & 3.23 $\pm$ 1.71 & 10 \\
NGC 2251 & $-$0.09 &   & 1 &   &   &   & $-$0.08 &   & 0.5 & 2 & 9.26 $\pm$ 0.11 & 0.25 $\pm$ 0.14 & 6 \\
NGC 2264 & $-$0.13 &   & 1 &   &   &   &   &   &   &   & 8.66 $\pm$ 0.13 & 0.01 $\pm$ 0.01 & 16 \\
NGC 2266 & $-$0.44 &   & 1 &   &   &   & $-$0.38 &   & 0.5 & 4 & 11.11 $\pm$ 0.29 & 0.89 $\pm$ 0.35 & 12 \\
NGC 2287 & $-$0.11 & 0.01 & 2 &   &   &   & $-$0.11 & 0.09 & 3 & 1,2,5 & 8.45 $\pm$ 0.03 & 0.17 $\pm$ 0.09 & 11 \\
NGC 2301 &   &   &   &   &   &   & +0.05 & 0.01 & 1.5 & 2,5 & 8.64 $\pm$ 0.18 & 0.14 $\pm$ 0.08 & 11 \\
NGC 2324 & $-$0.22 & 0.07 & 2 &   &   &   & $-$0.08 & 0.21 & 2.5 & 1,3 & 11.20 $\pm$ 0.36 & 0.54 $\pm$ 0.26 & 8 \\
NGC 2335 & $-$0.18 &   & 1 &   &   &   & $-$0.03 &   & 0.5 & 2 & 9.00 $\pm$ 0.20 & 0.15 $\pm$ 0.03 & 6 \\
NGC 2343 &   &   &   &   &   &   & +0.03 &   & 1.5 & 1 & 8.72 $\pm$ 0.07 & 0.10 $\pm$ 0.06 & 8 \\
NGC 2354 & $-$0.18 & 0.02 & 2 &   &   &   & $-$0.20 &   & 1 & 2 & 10.33 $\pm$ 0.39 & 0.20 $\pm$ 0.14 & 4 \\
NGC 2355 & $-$0.05 & 0.08 & 3 & $-$0.08 & 0.08 & 5 &   &   &   &   & 9.87 $\pm$ 0.19 & 0.81 $\pm$ 0.14 & 14 \\
NGC 2360 & $-$0.03 & 0.06 & 9 &   &   &   & $-$0.05 & 0.00 & 1.5 & 2,3 & 8.96 $\pm$ 0.29 & 1.10 $\pm$ 0.61 & 6 \\
NGC 2420 & $-$0.05 & 0.02 & 3 & $-$0.21 & 0.09 & 5 & $-$0.32 & 0.10 & 4 & 2,3,5,6 & 10.26 $\pm$ 0.34 & 2.34 $\pm$ 0.68 & 10 \\
NGC 2422 &   &   &   &   &   &   & +0.09 & 0.03 & 2.5 & 1,5 & 8.30 $\pm$ 0.02 & 0.10 $\pm$ 0.04 & 11 \\
NGC 2423 & +0.08 & 0.05 & 3 &   &   &   & +0.07 & 0.03 & 2.5 & 1,2 & 8.49 $\pm$ 0.06 & 0.79 $\pm$ 0.32 & 9 \\
NGC 2425 &   &   &   & $-$0.15 & 0.09 & 4 &   &   &   &   & 10.76 $\pm$ 0.39 & 2.21 $\pm$ 0.03 & 3 \\
NGC 2437 &   &   &   &   &   &   & $-$0.07 &   & 0.5 & 2 & 9.01 $\pm$ 0.10 & 0.23 $\pm$ 0.05 & 8 \\
NGC 2447 & $-$0.05 & 0.01 & 3 &   &   &   & $-$0.02 & 0.03 & 2.5 & 1,2 & 8.57 $\pm$ 0.02 & 0.41 $\pm$ 0.15 & 9 \\
NGC 2451A &   &   &   &   &   &   & $-$0.08 &   & 1.5 & 1 & 8.06 $\pm$ 0.00 & 0.05 $\pm$ 0.01 & 5 \\
NGC 2451B &   &   &   &   &   &   & +0.00 &   & 1.5 & 1 & 8.11 $\pm$ 0.01 & 0.05 $\pm$ 0.01 & 4 \\
NGC 2477 & +0.07 & 0.03 & 4 &   &   &   & +0.13 &   & 1 & 3 & 8.45 $\pm$ 0.07 & 0.85 $\pm$ 0.20 & 16 \\
NGC 2482 & $-$0.07 &   & 1 &   &   &   & +0.10 &   & 0.5 & 2 & 8.58 $\pm$ 0.19 & 0.37 $\pm$ 0.07 & 5 \\
NGC 2489 &   &   &   &   &   &   & +0.05 & 0.02 & 2 & 1,2 & 8.98 $\pm$ 0.57 & 0.20 $\pm$ 0.16 & 11 \\
NGC 2506 & $-$0.23 & 0.05 & 5 &   &   &   & $-$0.30 & 0.07 & 3.5 & 1,2,3 & 10.38 $\pm$ 0.19 & 1.63 $\pm$ 0.43 & 20 \\
NGC 2516 & +0.05 & 0.11 & 2 &   &   &   & +0.04 & 0.03 & 3 & 1,2,5 & 7.98 $\pm$ 0.00 & 0.13 $\pm$ 0.06 & 16 \\
NGC 2527 & $-$0.10 & 0.04 & 2 &   &   &   & $-$0.08 &   & 0.5 & 2 & 8.26 $\pm$ 0.02 & 0.65 $\pm$ 0.17 & 6 \\
NGC 2539 & $-$0.02 & 0.08 & 4 &   &   &   & +0.14 &   & 1 & 2 & 8.80 $\pm$ 0.09 & 0.50 $\pm$ 0.12 & 8 \\
NGC 2546 &   &   &   &   &   &   & +0.01 & 0.06 & 2 & 1,2 & 8.28 $\pm$ 0.06 & 0.09 $\pm$ 0.04 & 8 \\
NGC 2547 &   &   &   &   &   &   & $-$0.14 & 0.10 & 2.5 & 1,5 & 8.05 $\pm$ 0.01 & 0.05 $\pm$ 0.02 & 10 \\
NGC 2548 &   &   &   &   &   &   & +0.08 &   & 0.5 & 2 & 8.50 $\pm$ 0.04 & 0.38 $\pm$ 0.10 & 10 \\
NGC 2567 & $-$0.04 & 0.08 & 3 &   &   &   & $-$0.01 & 0.01 & 2 & 1,2 & 8.74 $\pm$ 0.06 & 0.30 $\pm$ 0.10 & 10 \\
NGC 2632 & +0.16 & 0.08 & 22 &   &   &   & +0.13 & 0.03 & 3 & 1,2,5 & 8.13 $\pm$ 0.01 & 0.73 $\pm$ 0.19 & 12 \\
NGC 2658 &   &   &   &   &   &   & $-$0.01 &   & 1.5 & 1 & 9.84 $\pm$ 0.23 & 0.25 $\pm$ 0.06 & 5 \\
NGC 2660 & +0.04 & 0.03 & 4 &   &   &   & $-$0.18 &   & 1 & 2 & 8.66 $\pm$ 0.06 & 1.31 $\pm$ 0.26 & 7 \\
NGC 2682 & +0.03 & 0.05 & 27 & +0.00 & 0.06 & 14 & $-$0.01 & 0.08 & 4 & 2,3,4,5 & 8.57 $\pm$ 0.04 & 3.45 $\pm$ 1.13 & 27 \\
NGC 2972 &   &   &   &   &   &   & $-$0.09 &   & 0.5 & 2 & 8.06 $\pm$ 0.07 & 0.31 $\pm$ 0.22 & 4 \\
NGC 3114 & +0.05 & 0.06 & 2 &   &   &   & +0.04 & 0.02 & 2.5 & 1,2 & 7.84 $\pm$ 0.00 & 0.13 $\pm$ 0.05 & 9 \\
NGC 3228 &   &   &   &   &   &   & +0.01 &   & 1.5 & 1 & 7.92 $\pm$ 0.00 & 0.15 $\pm$ 0.09 & 6 \\
NGC 3532 & +0.00 & 0.07 & 4 & +0.08 & 0.06 & 4 & $-$0.00 & 0.02 & 2.5 & 1,2 & 7.85 $\pm$ 0.01 & 0.30 $\pm$ 0.08 & 10 \\
NGC 3680 & $-$0.01 & 0.06 & 10 & $-$0.10 & 0.13 & 12 & $-$0.09 & 0.10 & 4 & 2,3,5,6 & 7.78 $\pm$ 0.03 & 1.80 $\pm$ 0.86 & 12 \\
NGC 3960 & $-$0.04 & 0.10 & 5 &   &   &   & $-$0.10 & 0.10 & 3.5 & 1,2,3 & 7.41 $\pm$ 0.04 & 0.97 $\pm$ 0.22 & 7 \\
NGC 4337 & +0.12 & 0.05 & 7 &   &   &   &   &   &   &   & 7.11 $\pm$ 0.07 & 1.75 $\pm$ 0.43 & 4 \\
NGC 4349 & $-$0.07 & 0.06 & 2 &   &   &   & $-$0.06 &   & 1 & 2 & 7.30 $\pm$ 0.15 & 0.33 $\pm$ 0.18 & 7 \\
NGC 4815 & +0.04 & 0.06 & 4 &   &   &   &   &   &   &   & 6.87 $\pm$ 0.05 & 0.40 $\pm$ 0.15 & 11 \\
NGC 5138 &   &   &   &   &   &   & +0.12 &   & 0.5 & 2 & 7.10 $\pm$ 0.12 & 0.10 $\pm$ 0.10 & 6 \\
NGC 5281 &   &   &   &   &   &   & $-$0.02 &   & 1.5 & 1 & 7.25 $\pm$ 0.09 & 0.05 $\pm$ 0.02 & 7 \\
NGC 5316 &   &   &   &   &   &   & +0.13 &   & 0.5 & 2 & 7.31 $\pm$ 0.12 & 0.16 $\pm$ 0.04 & 7 \\
NGC 5460 &   &   &   &   &   &   & +0.03 &   & 1.5 & 1, & 7.54 $\pm$ 0.04 & 0.17 $\pm$ 0.05 & 9 \\
NGC 5662 &   &   &   &   &   &   & +0.04 &   & 1.5 & 1, & 7.52 $\pm$ 0.04 & 0.09 $\pm$ 0.04 & 10 \\
NGC 5822 & +0.08 & 0.08 & 7 & +0.09 &   & 1 & $-$0.02 & 0.04 & 3.5 & 2,3,5,6 & 7.39 $\pm$ 0.05 & 0.89 $\pm$ 0.23 & 9 \\
NGC 5999 &   &   &   &   &   &   & $-$0.03 &   & 1.5 & 1 & 6.39 $\pm$ 0.18 & 0.40 $\pm$ 0.07 & 5 \\
NGC 6031 &   &   &   &   &   &   & +0.00 &   & 1.5 & 1 & 6.68 $\pm$ 0.24 & 0.18 $\pm$ 0.06 & 8 \\
NGC 6067 &   &   &   &   &   &   & +0.14 &   & 1 & 2 & 6.62 $\pm$ 0.16 & 0.10 $\pm$ 0.02 & 9 \\
NGC 6087 &   &   &   & +0.21 &   & 1 & +0.04 &   & 1.5 & 1 & 7.27 $\pm$ 0.04 & 0.08 $\pm$ 0.02 & 11 \\
NGC 6134 & +0.11 & 0.07 & 8 &   &   &   & +0.18 & 0.05 & 3.5 & 1,2,5 & 7.19 $\pm$ 0.14 & 0.91 $\pm$ 0.33 & 13 \\
NGC 6192 & +0.12 & 0.07 & 3 &   &   &   & +0.15 & 0.02 & 2 & 1,2 & 6.56 $\pm$ 0.17 & 0.12 $\pm$ 0.03 & 8 \\
NGC 6204 &   &   &   &   &   &   & +0.02 &   & 1.5 & 1 & 7.00 $\pm$ 0.11 & 0.11 $\pm$ 0.06 & 11 \\
NGC 6208 &   &   &   &   &   &   & $-$0.01 &   & 0.5 & 2 & 7.08 $\pm$ 0.14 & 1.43 $\pm$ 0.40 & 7 \\
NGC 6253 & +0.34 & 0.11 & 10 & +0.44 & 0.08 & 53 & +0.55 & 0.04 & 2 & 5,6 & 6.60 $\pm$ 0.10 & 3.91 $\pm$ 1.13 & 9 \\
NGC 6268 &   &   &   &   &   &   & +0.14 &   & 0.5 & 2 & 6.97 $\pm$ 0.03 & 0.17 $\pm$ 0.16 & 7 \\
NGC 6281 & +0.06 & 0.06 & 2 &   &   &   & +0.02 & 0.01 & 2 & 1,2 & 7.45 $\pm$ 0.12 & 0.26 $\pm$ 0.07 & 7 \\
NGC 6405 &   &   &   &   &   &   & +0.07 &   & 1.5 & 1 & 7.54 $\pm$ 0.04 & 0.08 $\pm$ 0.02 & 10 \\
NGC 6425 &   &   &   &   &   &   & +0.16 &   & 0.5 & 2 & 7.15 $\pm$ 0.11 & 0.17 $\pm$ 0.15 & 4 \\
NGC 6451 &   &   &   &   &   &   & +0.01 &   & 1.5 & 1 & 5.84 $\pm$ 0.24 & 0.17 $\pm$ 0.07 & 6 \\
NGC 6475 & +0.02 & 0.02 & 3 & +0.11 & 0.04 & 9 & +0.06 & 0.04 & 3 & 1,2,5 & 7.73 $\pm$ 0.03 & 0.22 $\pm$ 0.10 & 13 \\
NGC 6494 & $-$0.04 & 0.08 & 3 &   &   &   & +0.09 &   & 0.5 & 2 & 7.34 $\pm$ 0.08 & 0.33 $\pm$ 0.10 & 6 \\
NGC 6583 & +0.37 & 0.04 & 2 &   &   &   &   &   &   &   & 6.15 $\pm$ 0.18 & 1.00 $\pm$ 0.00 & 3 \\
NGC 6603 &   &   &   &   &   &   & +0.43 &   & 1 & 4 & 5.44 $\pm$ 0.77 & 0.39 $\pm$ 0.20 & 7 \\
NGC 6633 & $-$0.08 & 0.12 & 4 & $-$0.11 & 0.09 & 7 & +0.01 &   & 1 & 2 & 7.72 $\pm$ 0.03 & 0.52 $\pm$ 0.25 & 10 \\
NGC 6705 & +0.12 & 0.09 & 21 & +0.25 & 0.05 & 3 & +0.18 & 0.03 & 2.5 & 1,4 & 6.45 $\pm$ 0.23 & 0.19 $\pm$ 0.08 & 13 \\
NGC 6756 &   &   &   &   &   &   & +0.08 &   & 1.5 & 1 & 6.54 $\pm$ 0.37 & 0.13 $\pm$ 0.09 & 4 \\
NGC 6791 & +0.42 & 0.05 & 8 & +0.35 & 0.07 & 22 & +0.44 & 0.06 & 4 & 2,3,4,5,6 & 7.69 $\pm$ 0.14 & 7.00 $\pm$ 2.46 & 9 \\
NGC 6802 &   &   &   &   &   &   & +0.01 &   & 1.5 & 1 & 7.21 $\pm$ 0.20 & 0.87 $\pm$ 0.43 & 10 \\
NGC 6811 & +0.03 & 0.01 & 3 &   &   &   &   &   &   &   & 7.87 $\pm$ 0.01 & 0.72 $\pm$ 0.18 & 10 \\
NGC 6819 & +0.09 & 0.01 & 3 & $-$0.04 & 0.08 & 206 & +0.03 & 0.14 & 4 & 3,4,5,6 & 7.69 $\pm$ 0.00 & 2.11 $\pm$ 0.44 & 8 \\
NGC 6830 &   &   &   &   &   &   & +0.22 &   & 1.5 & 1 & 7.28 $\pm$ 0.06 & 0.09 $\pm$ 0.05 & 10 \\
NGC 6939 & +0.13 &   & 1 & +0.02 & 0.05 & 6 & $-$0.02 & 0.07 & 1.5 & 3,4 & 8.26 $\pm$ 0.07 & 1.60 $\pm$ 0.39 & 9 \\
NGC 6940 &   &   &   &   &   &   & +0.15 &   & 1 & 3 & 7.75 $\pm$ 0.02 & 0.95 $\pm$ 0.46 & 7 \\
NGC 7044 &   &   &   &   &   &   & +0.01 &   & 1 & 4 & 8.38 $\pm$ 0.04 & 1.73 $\pm$ 0.41 & 13 \\
NGC 7092 &   &   &   &   &   &   & +0.00 &   & 1.5 & 1 & 8.02 $\pm$ 0.00 & 0.31 $\pm$ 0.16 & 11 \\
NGC 7142 & +0.11 & 0.04 & 4 & +0.08 & 0.06 & 6 & +0.07 & 0.14 & 2 & 3,4 & 8.80 $\pm$ 0.21 & 3.64 $\pm$ 1.57 & 11 \\
NGC 7209 &   &   &   &   &   &   & +0.07 &   & 0.5 & 2 & 8.18 $\pm$ 0.04 & 0.35 $\pm$ 0.10 & 9 \\
NGC 7243 &   &   &   &   &   &   & +0.03 &   & 1.5 & 1 & 8.16 $\pm$ 0.01 & 0.10 $\pm$ 0.06 & 10 \\
NGC 7245 &   &   &   &   &   &   & $-$0.15 &   & 1 & 4 & 9.03 $\pm$ 0.39 & 0.36 $\pm$ 0.16 & 9 \\
NGC 7296 &   &   &   &   &   &   & $-$0.02 &   & 1.5 & 1 & 8.82 $\pm$ 0.16 & 0.29 $\pm$ 0.12 & 6 \\
NGC 7789 & +0.05 & 0.07 & 5 & +0.06 & 0.07 & 33 & $-$0.06 & 0.04 & 3 & 2,3,4 & 9.00 $\pm$ 0.12 & 1.53 $\pm$ 0.20 & 12 \\
Pismis 2 &   &   &   &   &   &   & +0.22 &   & 1 & 3 & 9.16 $\pm$ 0.34 & 1.47 $\pm$ 0.30 & 3 \\
Ruprecht 4 &   &   &   & $-$0.09 & 0.05 & 3 &   &   &   &   & 11.63 $\pm$ 0.44 & 1.03 $\pm$ 0.25 & 5 \\
Ruprecht 18 &   &   &   &   &   &   & $-$0.01 &   & 0.5 & 2 & 8.68 $\pm$ 0.19 & 0.12 $\pm$ 0.10 & 6 \\
Ruprecht 115 &   &   &   &   &   &   & +0.01 &   & 1.5 & 1 & 6.29 $\pm$ 0.20 & 0.65 $\pm$ 0.36 & 6 \\
Ruprecht 130 &   &   &   &   &   &   & +0.00 &   & 1.5 & 1 & 5.69 $\pm$ 0.70 & 0.26 $\pm$ 0.24 & 4 \\
Ruprecht 147 & +0.16 & 0.08 & 5 &   &   &   &   &   &   &   & 7.78 $\pm$ 0.06 & 2.37 $\pm$ 0.20 & 3 \\
Saurer 1 & $-$0.38 & 0.00 & 2 &   &   &   &   &   &   &   & 21.38 $\pm$ 1.74 & 3.49 $\pm$ 1.32 & 3 \\
Stock 2 &   &   &   &   &   &   & +0.17 & 0.18 & 2 & 1,2 & 8.23 $\pm$ 0.05 & 0.16 $\pm$ 0.06 & 6 \\
Tombaugh 2 &   &   &   & $-$0.30 & 0.07 & 20 & $-$0.31 &   & 1 & 3 & 15.39 $\pm$ 1.93 & 1.46 $\pm$ 0.45 & 7 \\
Trumpler 5 & $-$0.44 & 0.07 & 4 & $-$0.47 & 0.05 & 3 & $-$0.30 &   & 1 & 4 & 10.40 $\pm$ 0.65 & 4.33 $\pm$ 1.93 & 8 \\
Trumpler 10 &   &   &   &   &   &   & $-$0.12 & 0.06 & 2.5 & 1,5 & 8.06 $\pm$ 0.00 & 0.04 $\pm$ 0.01 & 10 \\
Trumpler 20 & +0.17 & 0.05 & 13 & +0.09 & 0.10 & 5 &   &   &   &   & 6.90 $\pm$ 0.06 & 1.36 $\pm$ 0.65 & 6 \\

\end{longtable}
\tablefoot{The HQS, LQS, and weighted photometric metallicities,  respective errors,  number of stars used for the calculation (N), and the total weight (W) of the photometric metallicities. The code for the photometric systems is 1 (DG), 2 (DDO), 3 (Friel), 4 (Ca II), 5 (Str\"omgren), 6 (Caby). The last column indicates the number of studies used to calculate the mean distance from the Sun and the age of the open clusters.}
\end{landscape}
}


\begin{acknowledgements}
MN acknowledges the support by the grant 14-26115P of the Czech Science Foundation. UH acknowledges support from the Swedish National Space Board (Rymdstyrelsen). EP is financed by the SoMoPro II programme (3SGA5916). The research leading to these results has acquired a financial grant from the People Programme (Marie Curie action) of the Seventh Framework Programme of EU according to the REA Grant Agreement No. 291782. The research is further co-financed by the South-Moravian Region. It was also supported by the grant 7AMB14AT015,
the financial contributions of the Austrian Agency for International 
Cooperation in Education and Research (BG-03/2013 and CZ-09/2014). 
This research has made use of the WEBDA database, operated at the Department of 
Theoretical Physics and Astrophysics of the Masaryk University. This work reflects only the author's views and the European Union is not liable for any use that may be made of the information contained therein. We thank Ivan Minchev for providing the Galactic chemical and chemodynamical models, and the referee, Bruce A. Twarog, for his useful comments and suggestions.
\end{acknowledgements}



\bibliography{26370} 
\bibliographystyle{aa} 


\end{document}